\documentclass[12pt,letter]{article}

\usepackage{graphicx, epsfig, color}
\usepackage{amsmath,amssymb,hyperref}
\usepackage{subfigure}
\usepackage[normalem]{ulem}

\def\eslt{\not\!\!{E_T}}
\def\to{\rightarrow}

\def\bi{\begin{itemize}}
\def\ei{\end{itemize}}

\def\sps1ap{SPS1a$^\prime$}
\def\c1p{C1$^\prime$}

\def\be{\begin{equation}}  
\def\ee{\end{equation}}  
\def\bea{\begin{eqnarray}}  
\def\eea{\end{eqnarray}}  
\def\beas{\begin{eqnarray*}}  
\def\eeas{\end{eqnarray*}}

\begin{document}
\begin{titlepage}
\begin{flushright}
UH-511-1304-19
\end{flushright}

\vspace{0.5cm}
\begin{center}
{\Large \bf Determining the space-time structure of bottom-quark couplings to spin-zero particles
}\\ 
\vspace{1.2cm} \renewcommand{\thefootnote}{\fnsymbol{footnote}}
{\large Tathagata Ghosh$^1$\footnote[1]{Email: tghosh@hawaii.edu }, 
Rohini Godbole$^2$\footnote[2]{Email: rohini@iisc.ac.in },
Xerxes Tata$^{1}$\footnote[3]{Email: tata@phys.hawaii.edu }
}\\ 
\vspace{1.2cm} \renewcommand{\thefootnote}{\arabic{footnote}}
{\it 
$^1$Department of Physics and Astronomy,
University of Hawaii, Honolulu, HI 96822, USA \\
}
{\it 
$^2$Centre for High Energy Physics, Indian Institute of Science,
Bangalore, 560012, India \\
}
\end{center}

\vspace{0.5cm}
\begin{abstract}
\noindent 

We present a general argument that highlights the difficulty of
determining the space-time structure of the renormalizable bottom quark
Yukawa interactions of the Standard Model Higgs boson, or for that
matter of any hypothetical spin-zero particle, at high energy
colliders. The essence of the argument is that, it is always possible,
by chiral rotations, to transform between scalar and pseudoscalar Yukawa
interactions without affecting the interactions of bottom quarks with SM
gauge bosons. Since these rotations affect only the $b$-quark mass terms
in the Standard Model Lagrangian, any differences in observables for
scalar versus pseudoscalar couplings vanish when $m_b\to 0$, and are
strongly suppressed in high energy processes involving the heavy
spin-zero particle where the $b$-quarks are typically relativistic. We
show, however, that the energy dependence of, for instance, $e^+e^- \to
b\bar{b} X$ (here $X$ denotes the spin-zero particle) close to the
reaction threshold may serve to provide a distinction
between the scalar versus pseudoscalar coupling at electron-positron
colliders 
{that are being proposed,} provided that the
$Xb\bar{b}$ coupling is sizeable. {We {also} note that while various kinematic distributions for $t \bar{t} h$ 
 are indeed sensitive to the space-time structure of the top Yukawa coupling,  for a spin-0  particle {$X$}
 of an arbitrary mass, the said sensitivity is lost if $m_{X} >> m_t$.}

\end{abstract}

\end{titlepage}

\section{Introduction}
\label{sec:intro}

 {With the discovery of the spin-zero Higgs-like boson $h$ in 2012~\cite{lhc_h} at the CERN LHC, the entire particle content of the Standard Model (SM)  has been experimentally observed.}
That this new particle, plays a role in electroweak symmetry
breaking is clear from the existence of $hZZ$ and $hW^+W^-$ couplings
whose space-time structure and size are close to SM expectations
\cite{hVV}, with only (loop-level) suppressed deviations from their SM
values possible. All measurements of the properties of $h$, to date, are
compatible with $h$ being the SM Higgs boson. These include measurements
of couplings of $h$ to third generation fermions, tau \cite{lhctau},
bottom \cite{lhcbottom} and top \cite{lhctop}, and an upper limit on the
muon Yukawa coupling just a factor of 2-3 above its SM expectation
\cite{cmsmuon}.

We stress that while the properties of $h$ are perfectly compatible with
those of the SM Higgs boson, the space-time structure of its couplings
to fermions 
is as yet
very poorly determined.  The possibility that $CP$ violation (CPV) may
be present in $hf\bar{f}$ {couplings} at tree-level, via the interaction terms,
\be
{\cal{L}}_{int} = y_f\bar{f}\left(\cos\alpha_f+
i\sin\alpha_f\gamma_5\right)f h,
\label{eq:cpvlag}
\ee
is not experimentally excluded. {Indeed CPV interactions of spin-zero particles with fermions could arise
in models where $h$ comprises of both $CP$ even and $CP$ odd components.} In contrast, $CP$-violating $hVV$ ($V=W,Z$) interactions 
{can arise only at the loop level as long as  CP is not violated spontaneously and will reflect themselves as non-renormalisable, higher dimensional operators.}  

LHC constraints from the measurements of various Higgs boson decay rates
on CPV top quark Yukawa interactions have been examined in
Ref.\cite{fawzi}. These authors, and many others
\cite{Ellis:2013yxa,Frederix:2011zi,Demartin:2014fia,Santos:2015dja,AmorDosSantos:2017ayi,Heinemeyer:2013tqa,Biswas:2014hwa,gunionhe,gritsan} analyse various kinematic distributions, 
spin correlations and CP-violating observables that would be possible to
measure at the LHC and show that these could be used to constrain
$\alpha_t$ in Eq.~(\ref{eq:cpvlag}).  Indeed, the space-time structure
of the top Yukawa interaction may also be probed via studies of top
polarization and CPV asymmetries of top decay products in $e^+e^- \to
t\bar{t}h$ at electron-positron colliders \cite{gunion,rohiniee}.  Many
groups \cite{hantau} have suggested $h\to \tau^+\tau^-$ decays and
constructed CPV asymmetries out of $\tau^{\pm}$ polarization vectors (some directly observable,
others proxies for observable quantities) that may be used to restrict
the value of $\alpha_{\tau}$ at the LHC.

While considerable attention has been devoted to the structure of the
top and tau Yukawa couplings, the bottom Yukawa has received only
limited attention \cite{gunionhe,gritsan}. Gunion and He \cite{gunionhe} 
note that an explicit evaluation of the $f\bar{f}h$ production from the
$gg$ or $q\bar{q}$ initial state shows that the $\alpha_f$ dependent
terms are proportional to $m_f^2$ (not counting any $m_f$ factors in the
Yukawa coupling), and so  significant only for $f=t$
(since for other quarks the $m_f^2$ term is negligible compared with
typical dot products of four-momenta in the process) which they then focus
on. Gritsan {\it et al.} \cite{gritsan} examine several parton-level kinematic
distributions for the $bbh$ final state and confirm that there is
essentially no detectable dependence on $\alpha_b$. 

Initially, our goal was to examine {the
prospects for determining} the space-time structure of the
(renormalizable) bottom quark Yukawa couplings of the Higgs-like boson
via measurements at high energy colliders such as the LHC or a future
electron-positron Higgs boson factory. During the course of our study,
we arrived at an understanding (based on chirality arguments presented
in Sec.~\ref{sec:chiral}) as to {\em
  why} earlier studies \cite{gunionhe,gritsan} found no observable
dependence on $\alpha_b$. We find that while the small Yukawa coupling
certainly makes things more difficult, the real underlying reason is the
smallness of the $b$-mass relative to the energy scale of the {hard scattering} process
(set by $m_h$). Put differently, our arguments clearly illustrate the
issues with determining the space-time structure of the Yukawa
interaction of a spin-zero hypothetical particle $X$ to relatively light
fermions, {\em even if the associated coupling is order unity.}  While
our conclusions about prospects for determining the space-time structure
of the Yukawa coupling {of the bottom quark} are largely negative (see Sec.~\ref{sec:thresh}
for an exception to the general arguments), and only confirm the
findings via explicit computations in the literature, we felt that the
deeper understanding that we have gained about the underlying reason for
this is worthwhile to report.

The chirality arguments that lead us to conclude that no physics can
depend on $\alpha_f$ are valid in the limit $m_f \to 0$, and break down
when fermion mass effects are important, {{\it i.e.} when the fermion is non-relativistic, as is the case close to
the kinematic threshold for the production process.} 
With this in mind, we also examine the threshold behaviour of
$2\to 3$ $Xb\bar{b}$ processes {with a large value of $ X b \bar{b} $ coupling} in Sec.~\ref{sec:thresh}. 
{These results, which we believe to be new,}
 offer an in-principle way of distinguishing $\alpha_f=0$ from $\alpha_f=\pi/2$ at, for example, an $e^+e^-$
collider operating just above the energy threshold for $b\bar{b}X$
production.

The remainder of this paper is organized as follows. In
Sec.~\ref{sec:chiral} we present our argument based on chirality to
demonstrate that the angle $\alpha_f$ becomes unobservable in the limit
that the fermion mass vanishes. This then provides a dynamical
understanding of the results in Ref. \cite{gunionhe,gritsan} where
explicit computations showed that the effects of $\alpha_b$ in
$hb\bar{b}$ production at the LHC are too small to be observed, even
though corresponding studies of $t\bar{t}h$ production illustrate
techniques for the determination of $\alpha_t$. Our argument also
illustrates why 
{it will not be possible to extract $\alpha_b$ using {\em kinematic variables in any process} at high energy facilities.}  
  In the next section, we provide explicit illustrations of the chirality arguments of
Sec.~\ref{sec:chiral}, for both the SM Higgs boson and the
hypothetical spin-zero $X$-boson with large Yukawa couplings mentioned
above. In Sec.~\ref{sec:thresh} we derive the threshold behaviour of
$2\to 3$ processes and apply our results to $b\bar{b}X$ production
processes, where $X$ couples to the $b\bar{b}$ pair as in
Eq.~(\ref{eq:cpvlag}), either with $\alpha_f=0$ or with
$\alpha_f=\pi/2$. 
{In Sec.~\ref{sec:top} we digress from the main theme of this paper. We 
note that while the various kinematic distributions in $t \bar t h$ are
indeed sensitive to the space time structure of the top Yukawa coupling,  the same is not
true for a spin-0  particle {$X$},  of an arbitrary mass.  The said sensitivity is lost for 
$m_{X} >> m_t$.}
We end in Sec.~\ref{sec:concl} with a summary of our results and
some general conclusions.

\section{Chirality Considerations and $CP$ violation} \label{sec:chiral}

We begin by noting that the Lagrangian in Eq.~(\ref{eq:cpvlag}) can be
rewritten after chiral
transformations, $$b_{R,L}={e^{i\theta^b_{R,L}}}b'_{L,R}$$ as
\be
{\cal L}_{\rm int} = \overline{b'_R}
e^{-i(\theta^b_R-\theta^b_L)} e^{-i\alpha_b} b'_L h+ h.c.
\ee
We thus see that by choosing $\theta^b_L-\theta^b_R = \alpha_b$, we can
rotate away the $CP$-violating term in the Lagrangian in
Eq.~(\ref{eq:cpvlag}).  Of course, such a chiral transformation would
lead to a pseudoscalar bilinear term proportional to $m_b$, and we would
have achieved nothing. However, if $m_b=0$, there would be no such
{term}. It would then appear that any $CP$ violating
  effects from the Lagrangian in Eq.~(\ref{eq:cpvlag}) must vanish if
  $m_b=0$. 

Note that the charged and neutral current couplings to the vector bosons
of the SM are also unaltered, as long as we perform a common chiral
transformation for the electroweak doublet, {\it i.e.} we take
$\theta^t_L=\theta^b_L \equiv \theta_L$ (as we must to preserve
$SU(2)_L$). We still have the freedom to make arbitrary chiral
transformations on $t_R$, since there is no right-handed charged vector
current in the SM. We can then use this freedom and choose
$\theta^t_R=\theta_L$ to keep the top quark mass term in the standard
form. Any contribution of the form of (\ref{eq:cpvlag}) to the top quark
sector would be left unaltered. If the top quark Yukawa has the SM form
({\it i.e.} $\alpha_t=0$), any effect of the space-time structure of the
bottom quark Yukawa coupling must {vanish} as $m_b\to 0$.  In particular,
any $\alpha_b$ dependence in $tbh$ production processes vanish with the
$b$-mass.

We thus conclude that in the SM, {\em any $CP$ violating effects from
  the Lagrangian in Eq.~(\ref{eq:cpvlag}) with $f=b$ must vanish if
  $m_b=0$.} This observation, which may well be known to some
aficionados, provides a clear explanation of the negative
results~\cite{gunionhe,gritsan} about the prospects to observe
$\sin\alpha_b$ effects in bottom Higgs Yukawa couplings, even though
corresponding effects are readily observable in the top Yukawa sector~\cite{fawzi,Ellis:2013yxa,Frederix:2011zi,Demartin:2014fia,Santos:2015dja,AmorDosSantos:2017ayi,Heinemeyer:2013tqa,Biswas:2014hwa,gunionhe,gritsan}. Indeed,
because our argument is made at the Lagrangian level, it applies not
only to these processes, but to {essentially}{\em any process} that may be envisioned
to study $CP$ violating effects from bottom quark Higgs boson Yukawa
interactions. Moreover, since we did not use any properties of $h$
(other than its spin) in arriving at our result, our argument also
applies to the corresponding couplings of any spin-zero neutral particle
$X$ to SM fermions, at least for processes that do not simultaneously
involve also the SM Higgs boson Yukawa coupling.  We will refer to this
as {\em chirality protection} of $CP$ invariance in the Yukawa sector.

\section{Illustration of chirality protection} \label{sec:illus}

\subsection{The Standard Model Higgs particle} \label{subsec:sm}

Prospects for the exploration
of the spacetime structure of SM top interactions, both at the LHC and
at electron-positron colliders, have been examined {by many authors~\cite{fawzi,Ellis:2013yxa,Frederix:2011zi,Demartin:2014fia,
Santos:2015dja,AmorDosSantos:2017ayi,Heinemeyer:2013tqa,
Biswas:2014hwa,gunionhe,gritsan,gunion,rohiniee,Djouadi:2005gi,
Ananthanarayan:2013cia,Hagiwara:2016rdv}}. 
Many potential observables have been suggested for the determination of
$\alpha_t$. These observables naturally divide up into kinematic
quantities such as transverse momenta, angles, or invariant masses, and
polarization observables which depend on the fact that the top quark decays very rapidly {(compared to the hadronization time)} so that polarization information is maintained by its
decay products.  In contrast, the bottom quark typically hadronizes to
excited $b$-hadrons, which de-excite to lighter $b$-hadrons whose decay
rates to yet lighter states are (dynamically and/or kinematically)
suppressed and so may compete with the $b$-quark spin flip rate. As a
result, information of the $b$-polarization in the hard production
process is largely screened from the final bottom meson decay products
\cite{peskin}.\footnote{It has been noted \cite{peskin,Galanti:2015pqa}
  that $\Lambda_b$ baryons partially preserve the original $b$-quark
  polarization. However, since the probability of $b$-quark
  fragmentation to baryons is just a few percent, we do not consider
  this further in our study.} Since $b$-quark Yukawa interactions are
the main subject of this paper, we focus our attention on kinematic
observables from this point on.

Clearly, {{along with  $h \rightarrow \tau^+ \tau^-$ decay,} $f\bar{f} h$  production
offers the best prospects for studying the space-time
structure of the Yukawa interactions of $h$ using kinematic
distributions.} Before turning to the
discussion of the bottom Yukawa coupling, we quickly review what has
been done for the much-more-accessible, and therefore, more studied top
Yukawa interaction.  There are numerous studies of {$t\bar{t} h$
process} at the LHC. 
{Kinematic variables, many at parton level, have been constructed using the momenta of the $t,\bar{ t}$ and $h$ and hence can be constructed in laboratory frame. Various angular observables from the momenta of the $t$ decay products have also been constructed, both in the laboratory frame as well as some special frames such as the $t \bar t$ rest frame, as they carry information about spin-spin correlation between the $t$ and $\bar{ t}$ which in turn  is affected by the value of $\alpha_t$.  All these observables while being CP even, have the potential to distinguish between $\alpha_t=0$ and $\alpha_t = \pi/2$. These include:
\begin{itemize}
\item $M_{t \bar{t} h}, \  p_{T_h}, \ \Delta \phi(t,
  \bar{t})${~\cite{fawzi,Ellis:2013yxa,Heinemeyer:2013tqa,gritsan}}.

\item  $M_{t \bar{t}}, \ M_{t h}$~\cite{Ellis:2013yxa}.

\item $p_{T_h}, p_{T_t}, M_{t \bar{t}}$~\cite{Frederix:2011zi}

\item {{Pseudorapidity differences between the (anti)leptons 
{or bottom quarks} coming from the $ t$ and $\bar t$ decay respectively}, for a
$h$ with high transverse momentum:
$\Delta \eta_{l^+l^-}, \Delta\eta_{b \bar b}$~\cite{Demartin:2014fia}.}
  
 \item 
 The {lab-frame} angle $\Delta \theta^{lh}(l^+l^-)$ between $l^+$ and $l^-$ projected onto a plane perpendicular to the $h$ direction~\cite{fawzi}.
 
\item  $\Delta \phi ^{t,\bar t} (l^+,l^-)$ : Difference between the azimuthal angles of the $l^+$ in rest frame of  $t$ and
that of $l^-$ in the rest frame of $\bar t${~\cite{fawzi,Ellis:2013yxa,Demartin:2014fia,Heinemeyer:2013tqa,Biswas:2014hwa}}


\item $M_{t \bar{t} h}, \ \theta_t, \Phi^*_t, M_{t \bar{t}},
  \theta_h, \theta_b, \Phi_b$~\cite{gritsan}. Here, $\theta_t$ is the
  angle between the top quark direction and the opposite to Higgs
  direction in the $t \bar{t}$ frame; $\Phi^*_t$ is the angle between
  the decay planes of the $t \bar{t}$ system and $h \rightarrow f
  \bar{f}$ in the $t \bar{t} h$ frame; $\theta_b$ is the angle between
  the $b$ quark and opposite of the $W^+ W^-$ system in the $b \bar{b}$
  frame; and finally, $\Phi_b$ is the angle between the planes of the 
$b \bar{b}$ and $W^+ W^-$ systems in the $t \bar{t}$ frame. This paper
  also examines many other distributions, but we have picked out the
  ones that appear to give maximum distinction {between $\alpha_t=0$ and
  $\alpha_t=\pi/2$. 
  }
\end{itemize}
In addition, several $CP$-violating asymmetries~\cite{fawzi,Ellis:2013yxa,Santos:2015dja,AmorDosSantos:2017ayi} have been proposed for the purpose of extracting $\alpha_t$. Since these are not directly relevant to us, we do not list these here.} Very recently, Goncalves {\it et al.} \cite{Goncalves:2018agy} have
examined the structure of the top-Yukawa coupling using variables
related to $M_{T2}$ to aid in the selection of signal events.

\begin{figure}[!htp]
\centering
\includegraphics[height=2 in]{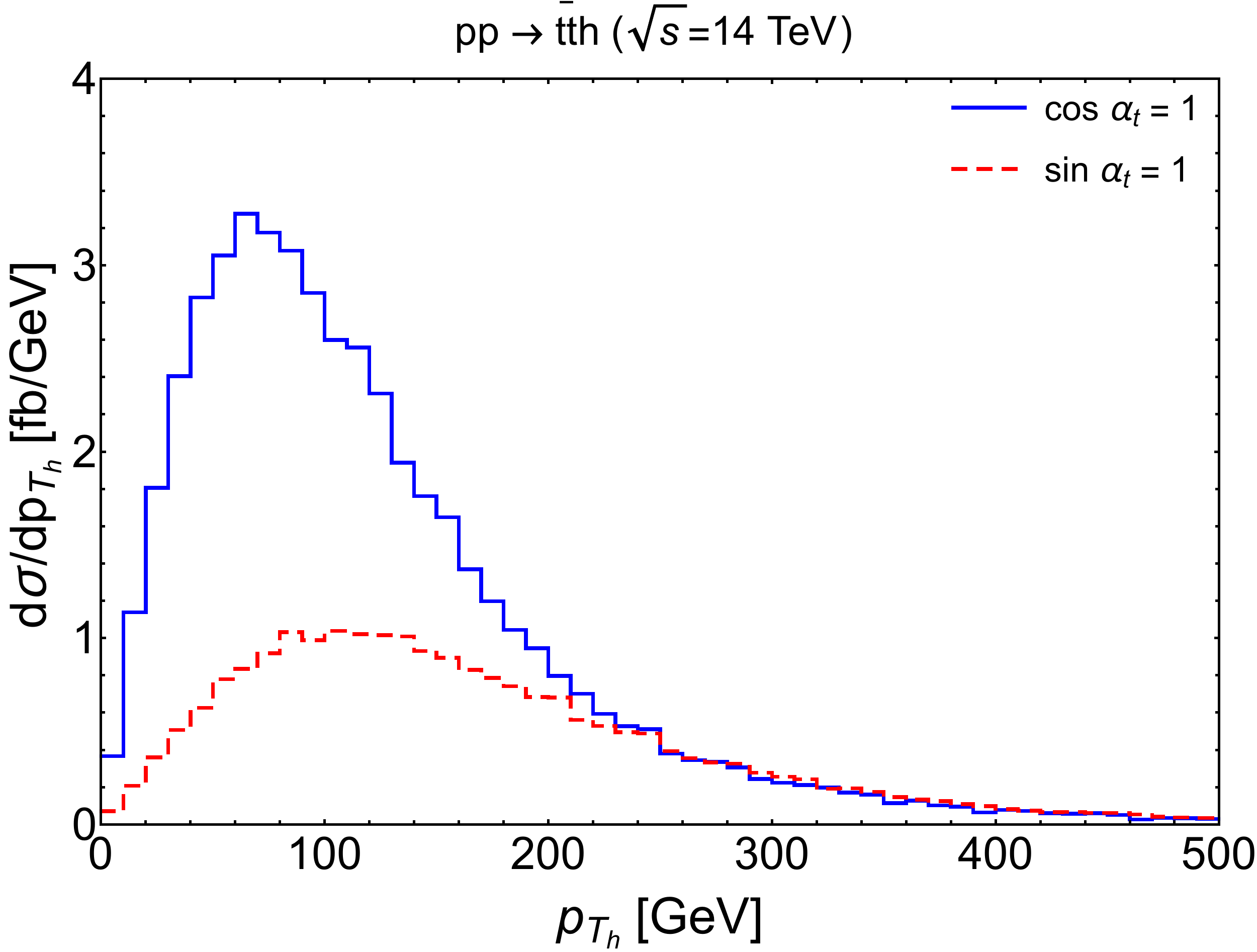}
\includegraphics[height=2 in]{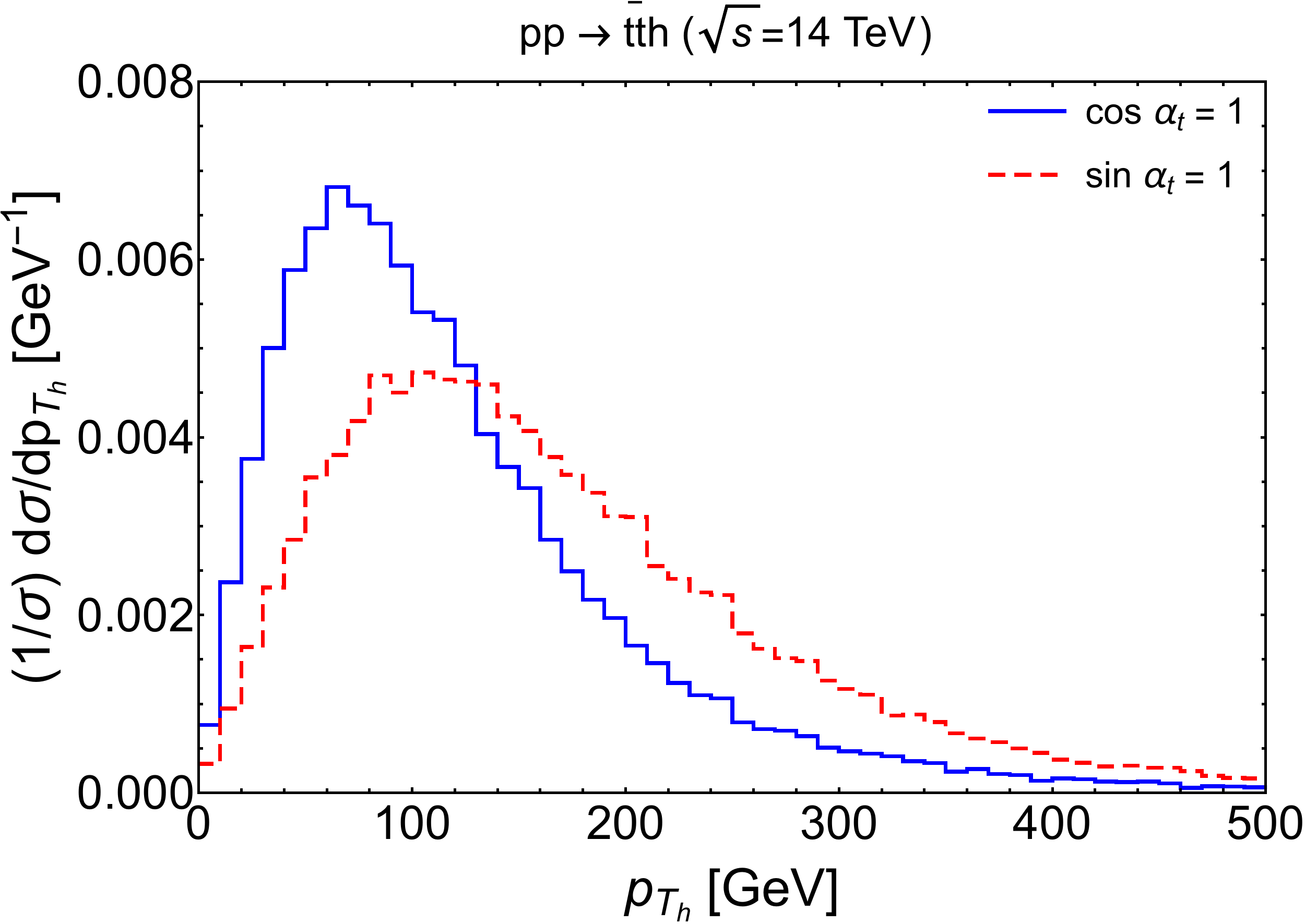}
\caption{[Left panel] The distribution of $p_{T_h}$ in $pp\to t
  \bar{t}h$ production at the LHC with $\sqrt{s}=14$~TeV, assuming that
  the Lagrangian Yukawa coupling is given by its SM value for both
  $\alpha_t=0$ as well as $\alpha_t= \pi/2$. [Right panel] The same
  distribution, normalized to unity.  The total cross-section for the
  two cases are 480.6 fb and 219.6 fb, respectively.
\label{fig:pth_LHC_tth}}
\end{figure}

To illustrate the degree of distinction between $\alpha_t=0$ and
$\alpha_t=\pi/2$ that may be possible at the LHC, we show in
Fig.~\ref{fig:pth_LHC_tth} the idealized $p_{Th}$ distribution for
$pp\to t\bar{t}h+X$ events, taking $m_h=125$~GeV. In the left panel, we
show the differential distribution assuming that the Yukawa coupling is
given by its SM value for both the scalar ($\alpha_t=0$, solid blue
line) and for the pseudoscalar ($\alpha_t=\pi/2$, dashed red line)
cases. The shapes, as well as the overall normalizations provide a clear
distinction between the two cases. Since the case $\alpha_t=\pi/2$ would
clearly be for a new particle with an unknown coupling, it is not
reasonable to use the absolute normalizations to distinguish between the
two cases. With this in mind, we show the corresponding distributions,
normalized to unity in the right-hand frames. We see that the shapes
alone provide a clear distinction, and so to the extent that it is
possible to reliably determine $p_{Th}$ in LHC events (this is not the
subject of this paper), it should be possible to distinguish between the
scalar and the pseudoscalar cases, and perhaps also obtain a measure of
$\alpha_t$. {We present these results here for completeness and 
have checked that they agree with the results available in the literature, for example, in Ref.~\cite{fawzi}.}

Prospects for examining the space-time structure of the top quark Yukawa
coupling {via $e^+e^- \to t\bar{t} h$ production} have also been examined
by numerous groups. The most promising ways include the dependence of
the total cross-section on the centre-of-mass energy
\cite{rohiniee,Ananthanarayan:2013cia}, a study of kinematic variables
$E_W, E_h$ \cite{Ananthanarayan:2013cia}, polarization aymmetries
\cite{rohiniee} and {toponium production in association with $h$}
\cite{Hagiwara:2016rdv}. We refer the interested reader to the
literature for details.

\begin{figure}[!htp]
\centering
\includegraphics[height=2 in]{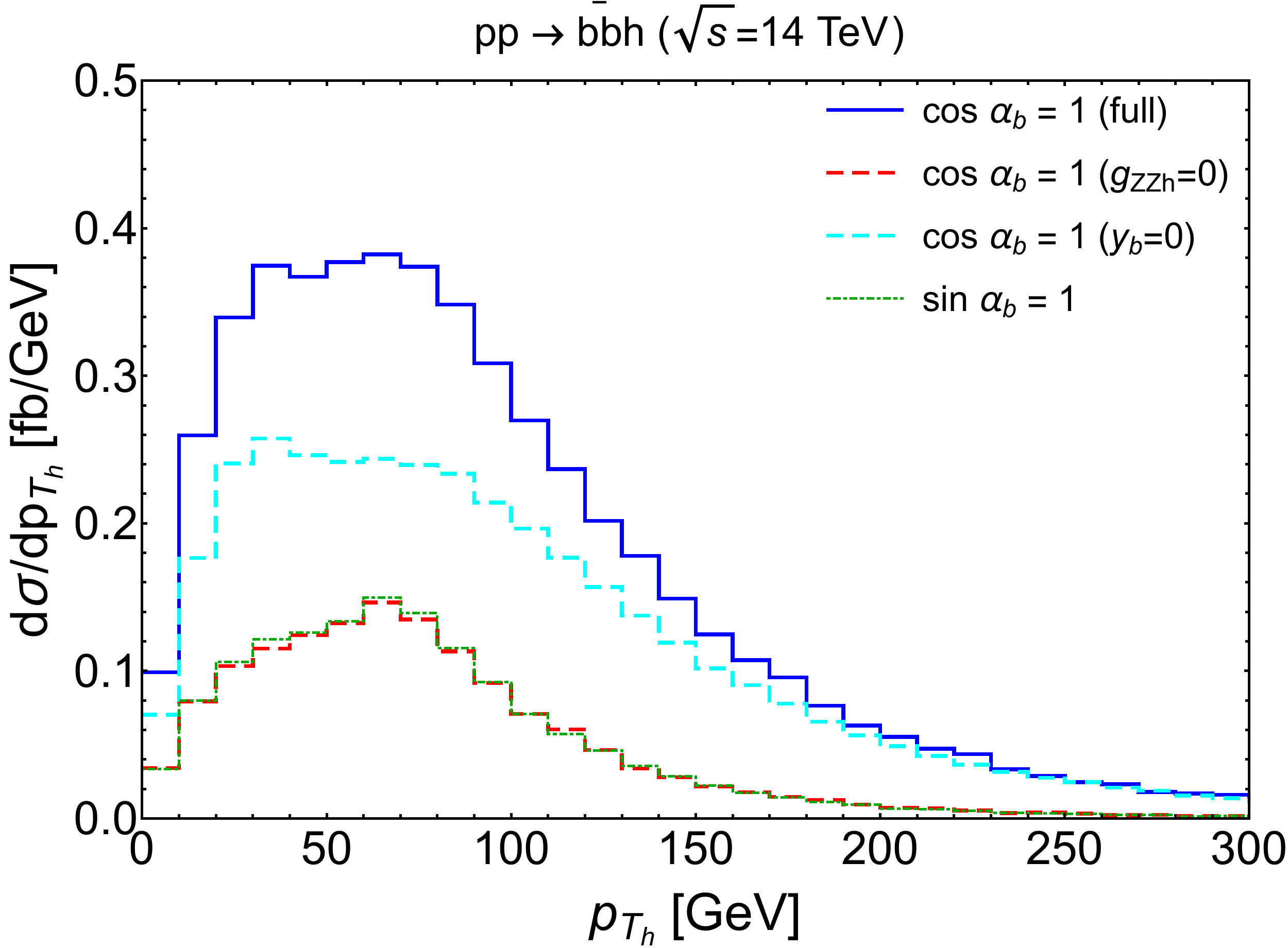}
\includegraphics[height=2 in]{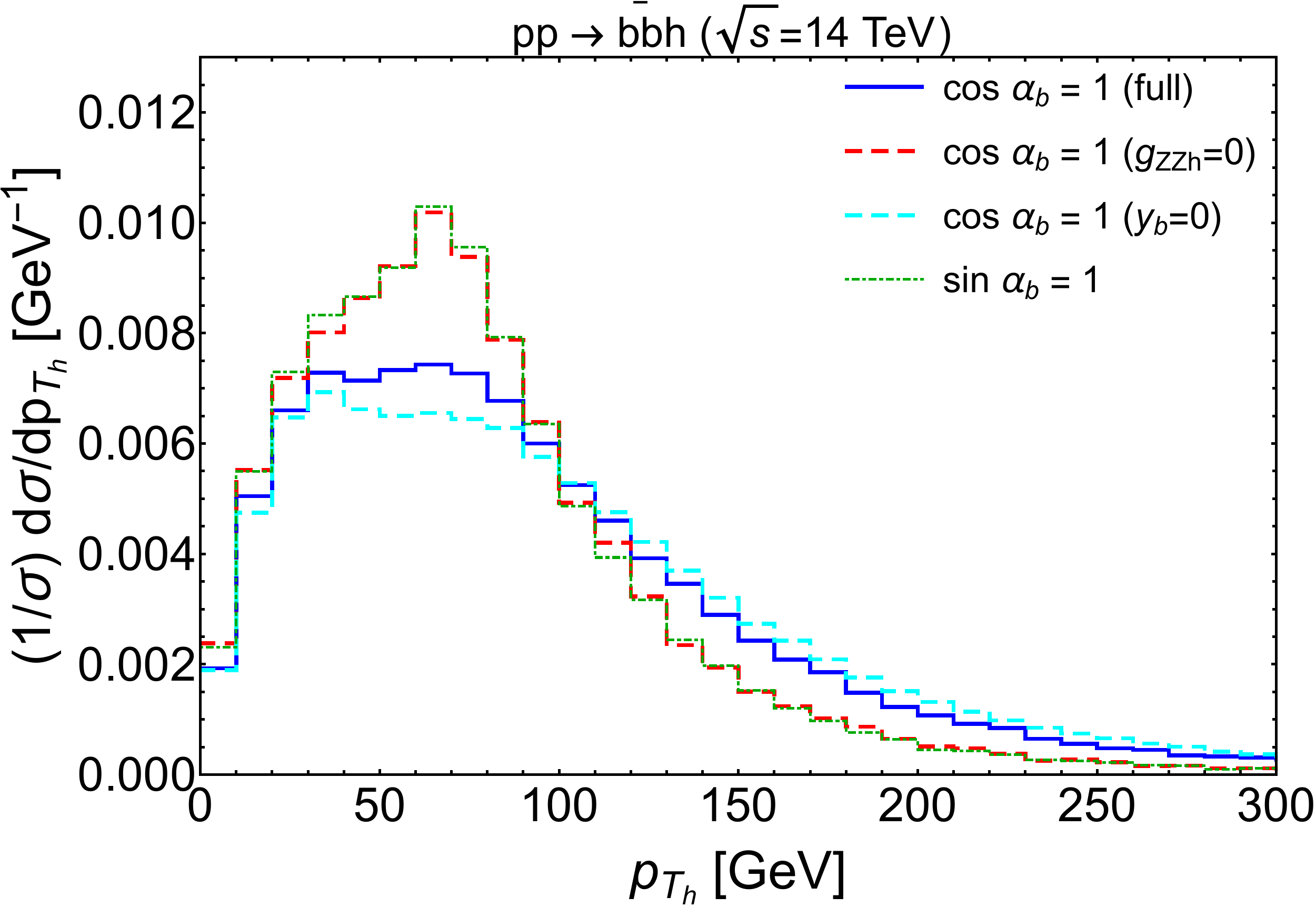}
\caption{[Left panel] The distribution of $p_{T_h}$ in $pp\to b
  \bar{b}h$ production at the LHC with $\sqrt{s}=14$~TeV, assuming that
  the Lagrangian Yukawa coupling is given by its SM value for both
  $\alpha_b=0$ as well as $\alpha_b= \pi/2$. [Right panel] The same
  distribution, normalized to unity. 
   The
  purely electroweak contribution where the $h$ is radiated off the
  virtual $Z$ boson is significant in this case. {After the $E_T(b) > 30$ GeV and $|\eta(b)| < 2.5$ cuts, the cross-section
  for the case $\cos \alpha_b = 1$ and $y_b = 0$  is 37.1~fb while the
corresponding total cross-section is 51.4~fb. The impact of the
  $ZZh$ coupling, which we assume is absent for the $\alpha_b=\pi/2$
  case, is illustrated by the three histograms shown.}
\label{fig:pth_LHC_bbh}}
\end{figure}

We begin our discussion of $pp \to b\bar{b}h$ at LHC14 by showing the
distribution of the transverse momentum of the $h$ parton in these
events in Fig.~\ref{fig:pth_LHC_bbh}, again assuming that the bottom
Yukawa coupling is given by its SM value.  We require $E_T(b)> 30$~GeV,
and $|\eta(b)| < 2.5$ to very roughly capture $b$-jet identification {effects}. 
We do not attempt to impose $b$-jet tagging efficiencies, so the cross
sections shown should be regarded as over-estimates. 
There is one significant
difference from the corresponding $pp \to t\bar{t}h$ production case
shown in the previous figure that is worth mentioning. Because of the
smallness of the bottom Yukawa coupling, electroweak production where
the $h$ is radiated off the (virtual) $Z$-boson now makes a {comparable contribution after $b$ quark $E_T$ and rapidity cuts}
for the case with $\alpha_b=0$~\footnote{{Without any $E_T(b)$ and $\eta(b)$ cuts the cross-sections for the $\cos \alpha_b =1, \, y_b=0$ case is 90.4 fb while the corresponding total coss-section is 946 fb. This is to be compared with 37.1 fb for $\cos \alpha_b =1, \, y_b=0$ case and a total cross-section of 51.4 fb that one obtains after cuts on $E_T(b)$ and $\eta(b)$.}}. Here, and in the remainder of this paper, we
assume that the couplings of the spin-zero particle to vector bosons
(which have been reasonably well determined) do not violate $CP$, so
that there is no $hZZ$ coupling when $\alpha_b=\pi/2$. As in
Fig~\ref{fig:pth_LHC_tth}, we show two frames, with the absolute cross
sections on the left, and cross sections normalized to unity on the
right. For the $\alpha_b=0$ case, we show three histograms to separate
out the pure electroweak contribution where the $h$ is radiated off the
$Z^*$, which is absent in the $\alpha_b=\pi/2$ case.

We see that there is a significant difference between the full $\alpha_b=0$ {(blue, solid)} and $\alpha_b=\pi/2$ (green, dot-dashed) histograms in both the left and right frames, suggesting that it is possible to distinguish between the two cases. We note, however that the differences are very sensitive to the details of the cuts that we impose on the $b$-parton. Given our very rudimentary treatment of $b$-jets these will also not reflect the {experimental $b$-jet} distributions. More importantly, these differences arise almost entirely from the $ZZh$ coupling, and do not reflect the difference between $\alpha_b=0$ and $\alpha_b = \pi/2$. Indeed we see that
the normalized
$\alpha_b=\pi/2$ (green, dot-dashed) histogram closely tracks the
corresponding $\alpha_b=0$ {with $ZZh$ coupling switched off} (red dashed) histogram in the right panel,
confirming that the differences of shape arise mostly from the
additional $ZZh$ coupling that is present for $\alpha_b=0$. This is an
explicit realization of our general result in Sec.~\ref{sec:chiral}
that any difference (other than due to the $hZZ$ coupling) between
$\alpha_b=0$ and $\alpha_b=\pi/2$ should vanish as $m_b \to 0$. Although
we have shown this for just the $p_{Th}$ distribution, we mention
in passing that similar results are obtained for other kinematic
variables.

Next, we turn to the examination of prospects for determining $\alpha_b$
at $e^+e^-$ colliders along the lines of corresponding studies for
$\alpha_t$ \cite{rohiniee,Ananthanarayan:2013cia}. Toward this end, we
consider $e^+e^-\to bbh$ {production at future electron-positron {colliders}, which {have been proposed} for precision studies}
of $Z$ and Higgs bosons, and of the $WW$ threshold
\cite{eecol}. This process occurs via amplitudes for $b\bar{b}$
production via a virtual $Z$, and where the $h$ is radiated off either
the quark line, or (for $\alpha_b \not= \pi/2$) off the $Z^*$.  We focus on operation at a centre-of-mass energy $\sqrt{s} = 161$~GeV, since the
process is kinematically inaccessible for $\sqrt{s}=M_Z$, and dominated 
by {by $2 \to 2$ $e^+e^- \to
 hZ(\rightarrow b \bar{b})$ production at the higher energy
 $\sqrt{s}=250$~GeV} 
envisioned for
detailed Higgs boson study. The integrated luminosity per interaction
region is evisioned to be $\sim 1.3$ (3.8)~ab$^{-1}$/yr for
the CEPC (FCC) design{~\cite{eecol}}.

\begin{figure}[!htp]
\centering
\includegraphics[height=2 in]{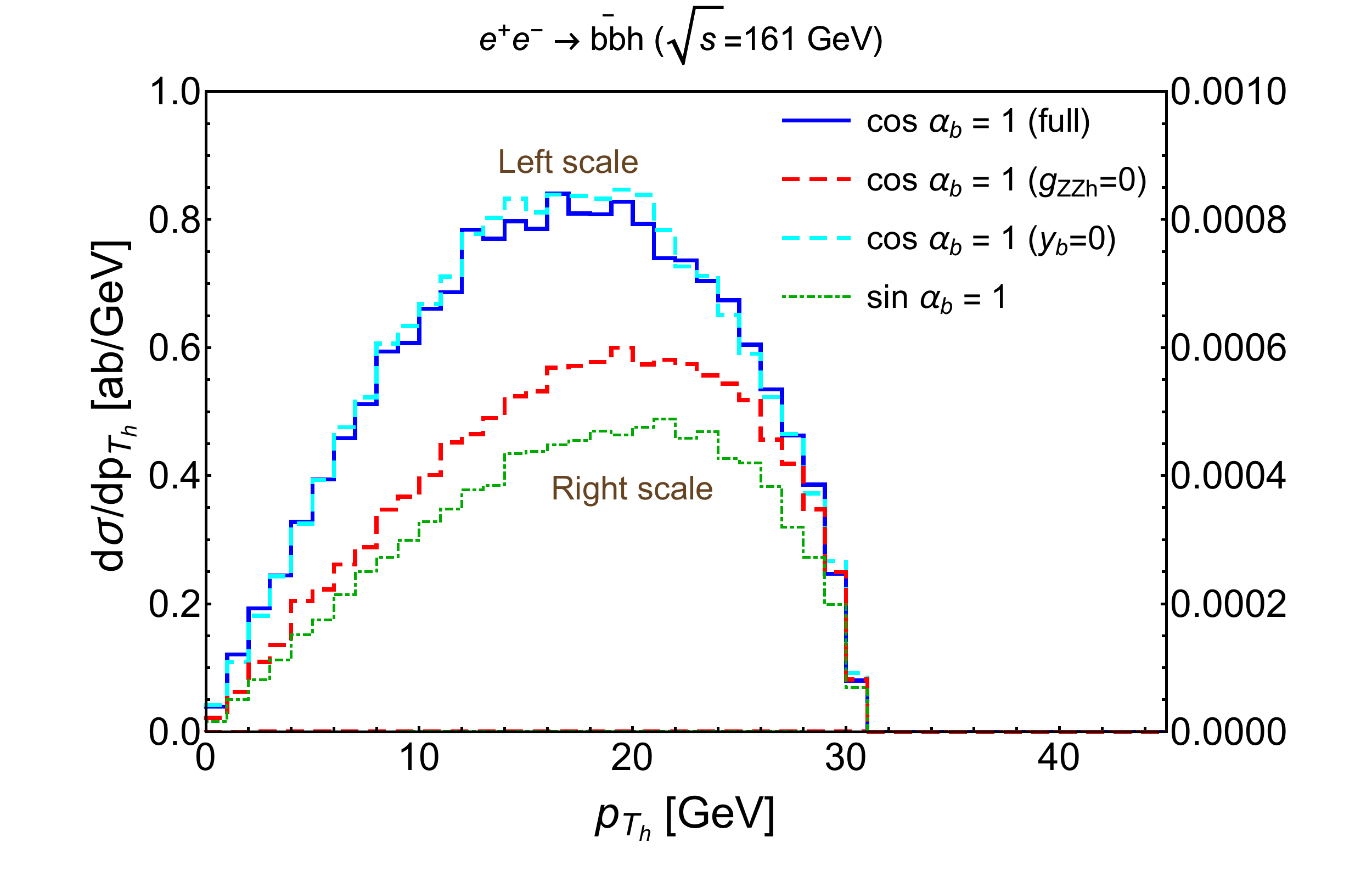}
\includegraphics[height=2 in]{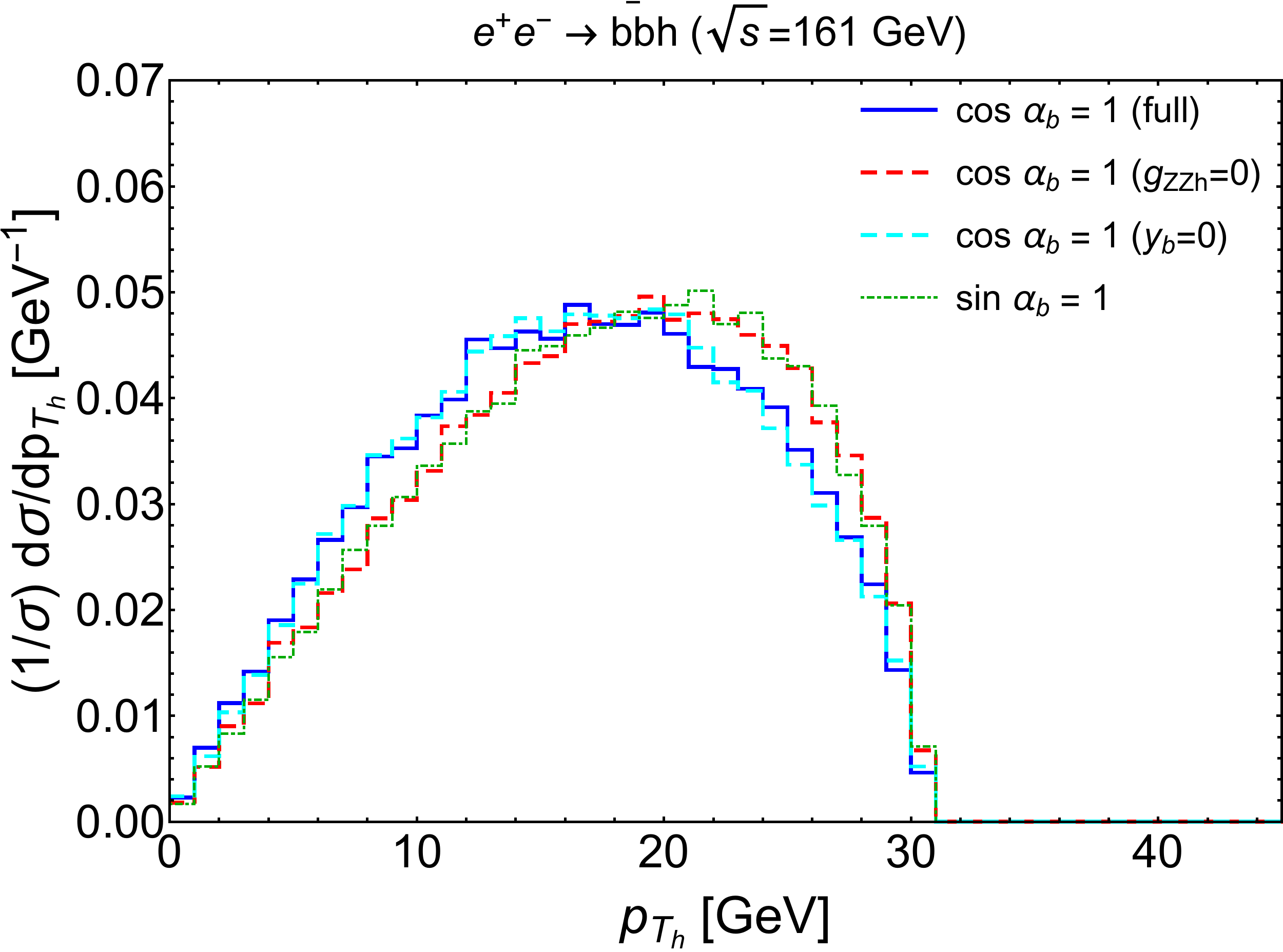}
\caption{The left panel shows the differential distribution of $p_{T_h}$
  produced via $e^+e^- \to b\bar{b} h$ at an electron-positron collider
  with $\sqrt{s}=161$~GeV for $\alpha_b=0$ and $\alpha_b=\pi/2$. As in
  Fig.~\ref{fig:pth_LHC_bbh}, we show three histograms for $\alpha_b=0$
  and one for $\alpha_b=\pi/2$. The dot-dashed green histogram for
  $\alpha_b=\pi/2$ and the dashed red histogram for $\alpha_b=0$ are to
  be read on the right-hand scale. The same distributions, but
  normalized to unity, are shown in the right panel. The total 
  cross-sections for $\cos \alpha_b =1$ and $\sin \alpha_b=1$ cases are
  $17.2$~ab and $9.75 \times 10^{-3}$~ab, respectively.  
\label{fig:pth_ILC_bbh}}
\end{figure}

As for the LHC studies just discussed, we show the $p_{Th}$ distribution
in Fig.~\ref{fig:pth_ILC_bbh}, again with the left (right) panels for
absolute values of the distributions (distributions normalized to
unity). As in Fig.~\ref{fig:pth_LHC_bbh} we assume that the $hZZ$
coupling vanishes if $\alpha_b=\pi/2$, and show three histograms for
$\alpha_b=0$ but just one histogram for $\alpha_b=\pi/2$. From the left
panels, we see instantly that the total cross section of 17.2~ab for
$\alpha_b=0$ is completely dominated by the $ZZh$ coupling, with per
mille size contributions $\simeq 9.75\times 10^{-3}$~ab (read off on the
right-hand scale) for $\alpha_b=\pi/2$. We conclude that the cross
section for $\alpha_b=\pi/2$ is unobservably small, and so there is no
chance of extracting the space-time structure of the bottom Higgs Yukawa
interaction or for that matter even testing for consistency with its SM
expectation.
Although this is only of academic interest, we see from the right panel
that the dashed red histogram for $\alpha_b=0$ with the $ZZh$ coupling
switched off essentially tracks the green, dot-dashed histogram for
$\alpha_b=\pi/2$. This is again exactly in keeping with what we would
expect from the chirality arguments of Sec.~\ref{sec:chiral}.  We
have verified this for several other kinematic distributions, but do not
show it for brevity. We again
conclude that the difference between the $\alpha_b=0$ and $\pi/2$ cases
arises (for example the green and the red lines in the left panel) {\em only} from the over-all normalization which is of no
practical use since the coupling of the pseudoscalar particle is not
known {\em a priori}.

For both, the LHC and an $e^+e^-$ collider, we saw that any
discernible differences between the scalar and pseudoscalar cases arose
only from the additional $ZZh$ coupling. The reader may well wonder
whether this occurred because the large gauge coupling of the SM Higgs
boson masked the tiny bottom Yukawa coupling. To assure ourselves that
this is not the essential reason, we turn to the examination of whether it is
possible to determine the space-time structure of the bottom quark
Yukawa couplings of a hypothetical spin-zero particle $X$ with sizeable
couplings to the $b$-quark but which does not develop a
vacuum-expectation-value and has no mixing with the Higgs boson, so
that the $ZZX$ coupling is absent.

\subsection{Yukawa interactions of a hypothetical scalar} \label{subsec:hyp}

We now turn to the prospects for determining the space-time structure of
the Yukawa interactions of a hypothetical spin-zero particle $X$ with
sizeable couplings to bottom quarks.  Since our main purpose here is
again to examine the impact of the chirality protection mechanism, for
definiteness we fix $m_X=100$~GeV, and allow $y_{bbX}$ to be as large as
possible. The most generic constraint on the Yukawa coupling of $X$
comes from the non-observation of an excess of events in a CMS search
for $X\to b\bar{b}$ events accompanied by at least one additional
$b$-jet at the 8~TeV LHC~\cite{Khachatryan:2015tra}. We note that there
are LHC13 analyses constraining the couplings of $X$, but these do not
extend down to $m_X=100$~GeV. For definiteness, we fix the Yukawa
coupling $y_{bbX}=0.7$, compatible with the CMS LHC8 constraints for
$m_X=100$~GeV~{\cite{Sirunyan:2018taj}}.\footnote{We mention that hypothetical spin-zero mediators
  that couple dark matter to SM fermions have been a subject of many  studies. In such scenarios, the $y_{bbX}$ coupling may be much more
  severely constrained using LHC data since invisible $X$ decays can lead to $\eslt$
  events \cite{buckley}. Such model-specific considerations are,
  however, irrelevant for our purposes.}
 
We focus our attention on $e^+e^-$ colliders where precision
measurements offer the best hope for distinguishing between a ``scalar''
($\alpha_b=0$) and {a} ``pseudoscalar'' ($\alpha_b=\pi/2$) $X$-boson. We
assume that the $XZZ$ {coupling} vanishes because $X$ does not develop a
vaccuum-expectation-value, and also does not mix with the SM Higgs
boson. We
consider $e^+e^- \to b\bar{b}X$ at $\sqrt{s}=161$~GeV, where we expect
essentially no physics background from SM $2\to 2$ processes, but
potentially important backgrounds from $Zb\bar{b}$ and $hb\bar{b}$
production, depending on the $m_{bb}$ mass resolution that is
attainable. The CERN Future Circular Collider (FCC) and the Circular
Electron Positron Collider (CEPC) being envisioned for construction
expect to accummulate 3.1~ab$^{-1}$/yr and 1.3~ab$^{-1}$/yr of
integrated luminosity, respectively, at each of two intersection
regions. Before any efficiency and acceptance considerations, we would
expect several hundred $e^+e^- \to b\bar{b}X$ per year at these
facilties for $y_{bbX} =0.7$. 
\begin{figure}[!htp]
\centering
\includegraphics[height=2 in]{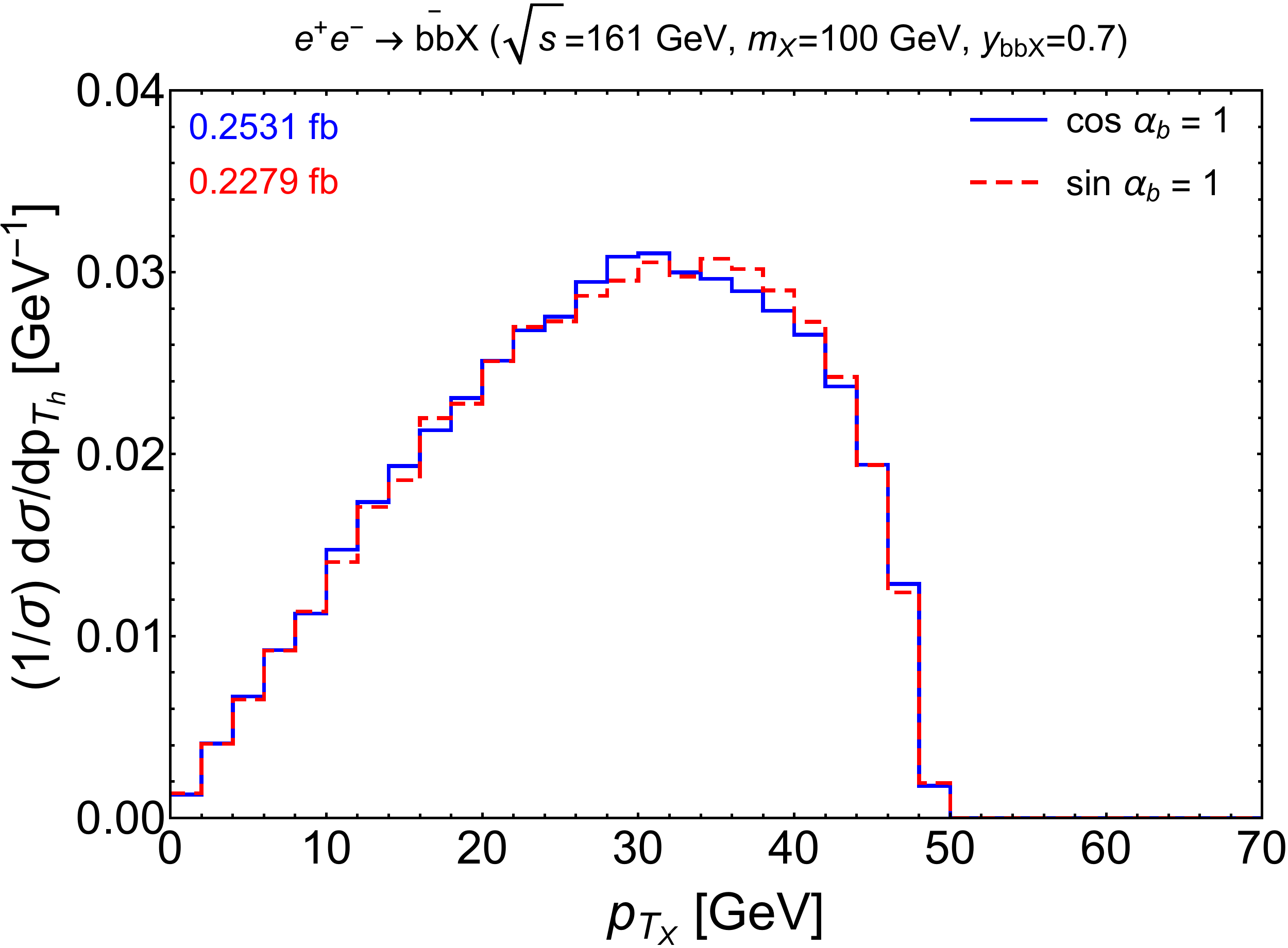}
\includegraphics[height=2 in]{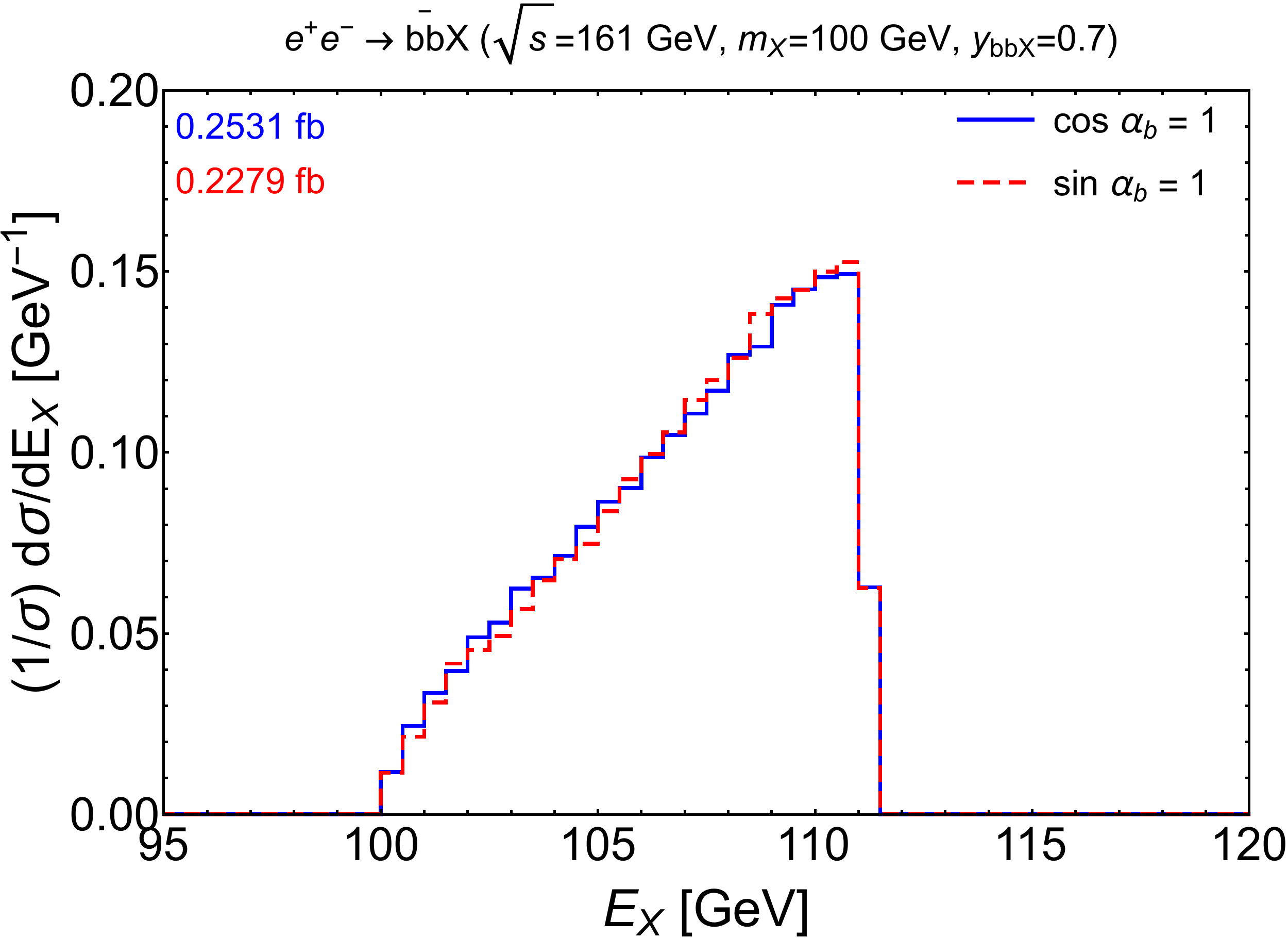}
\includegraphics[height=2 in]{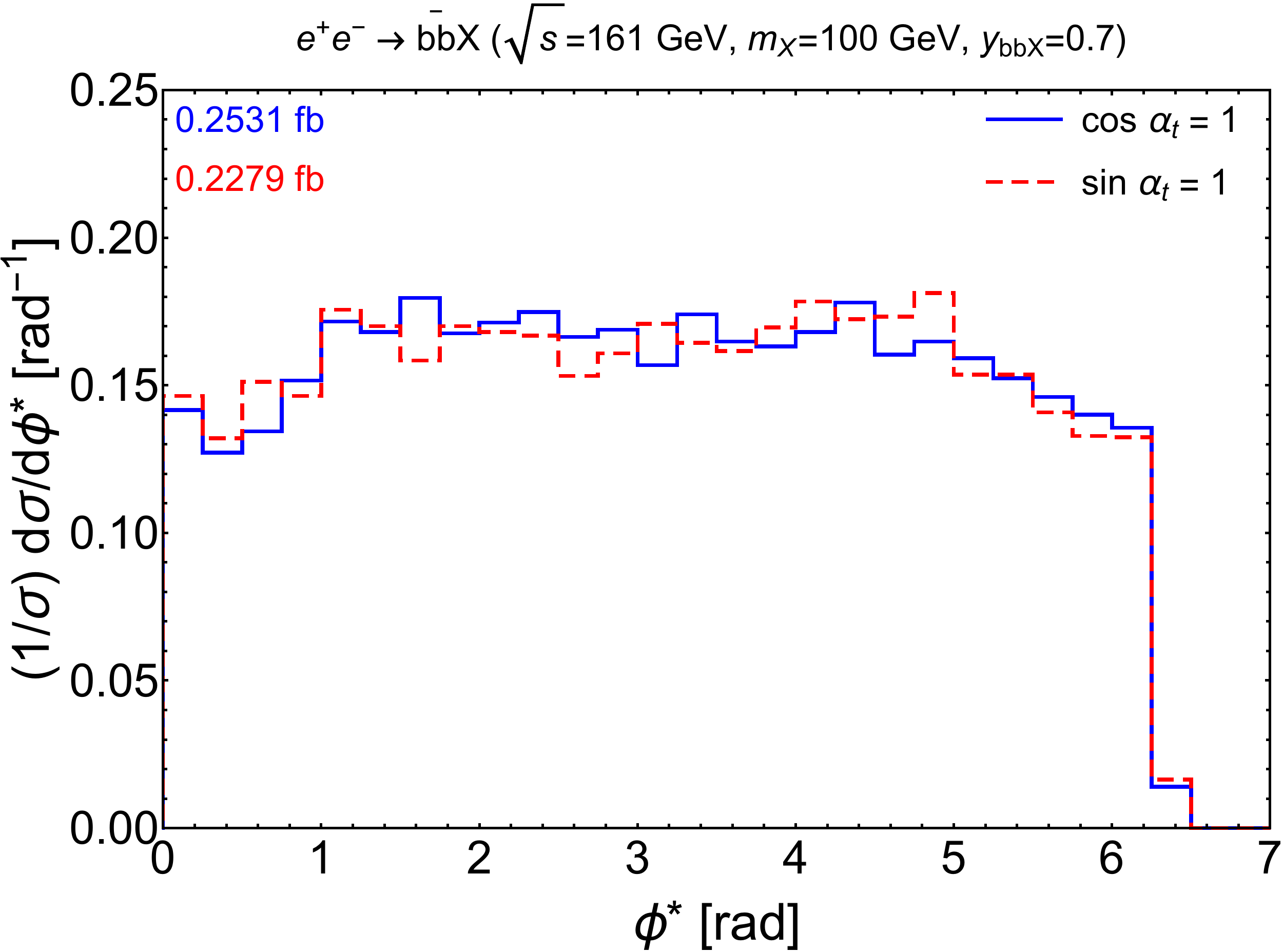}
\includegraphics[height=2 in]{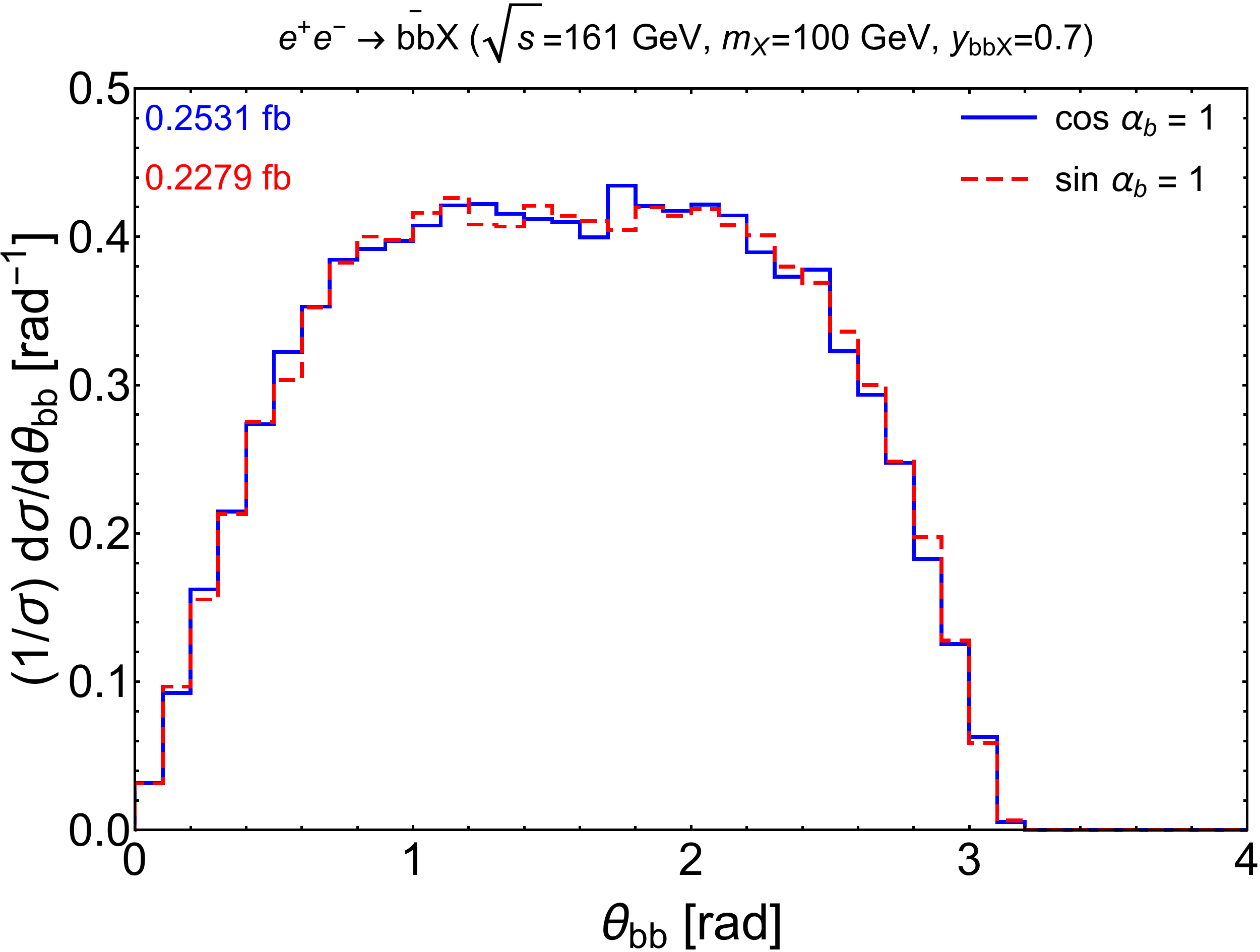}
\caption{Differential distributions for kinematic variables, $p_{T_X}$
  [upper left], $E_{X}$ [upper right], $\phi^*$ [lower left], $\theta_{b
    \bar{b}}$ [lower right] for the process $e^+e^-\to b\bar{b}X$ for
  (unpolarized initial beams) at a centre-of-mass energy $\sqrt{s}=161$
  GeV, where $X$ is a spin-zero particle.  All distributions are
  normalized to unity. We take $m_X=100$~GeV and the Yukawa coupling to
  bottom quarks $y_{bbX}=0.7$.
\label{fig:kin_x-ILC}}
\end{figure}

{We show parton-level results for distributions of various kinematic
observables from $e^+e^- \to b\bar{b}X$ production,} 
assuming unpolarized electron and positron beams, in
Fig.~\ref{fig:kin_x-ILC}. The solid, blue histogram is for $\alpha_b=0$
while the dashed, red histogram is for $\alpha_b=\pi/2$. We show results
only for the shapes of the distributions because the coupling $y_{bbX}$
which fixes the normalization 
is clearly  not known {\it a priori}.  Here,
$\theta_{bb}$ is the polar angle of the $b\bar{b}$ system in the
$e^+e^-$ centre-of-mass frame with respect to the direction of the
electron beam, while $\phi^*$ is the azimuthal angle of $b$ in the
$b\bar{b}$ rest frame. In other words, $\phi^*$ is the angle between the
plane formed by $e^+$, $e^-$ and $X$ and the plane formed by the $b$,
$\bar{b}$ and $X$, measured in the $b\bar{b}$ rest frame. We recognize
that, in practice, it will be difficult to know the $b$ and $\bar{b}$
directions (even if we can use kinematics to distinguish between the
primary bottom quarks from the secondary bottoms produced by decay of
the $X$). Our point, however, is that even if we ignore these practical
difficulties, we see from the figure that the distributions are
virtually identical (within expected statistics, even ignoring any
background) for $\alpha_b=0$ and $\alpha_b=\pi/2$.
This is, of
course, completely in keeping with what we would expect from the
chirality considerations of Sec.~\ref{sec:chiral}.


\section{Threshold Behaviour of $b\bar{b}X$ production} \label{sec:thresh}

The illustrative examples of the previous section serve to confirm that
the chirality protection mechanism of Sec.~\ref{sec:chiral} precludes
the possibility of determining $\alpha_b$ in high energy processes where
the typical process sub-energy is much larger than $m_b$, for both the
SM Higgs boson or for a hypothetical scalar with large couplings to the
bottom quark.  This naturally suggests that in order to obtain
sensitivity to $\alpha_b$, the kinematics needs to be such that
$b$-quark mass effects are not negligible; {\it i.e.} we are close to
the reaction threshold, where the reaction products are essentially
non-relativistic.

The threshold dependence of cross-sections for $2\to2$ reactions, it is
well-known, can be extracted using partial wave analysis because, in the
centre-of-mass system, the associated three-momentum is small so that the
contribution from the lowest partial wave dominates. A
similar analysis can be performed for the three-body final state, which
is characterized by two relative orbital angular momenta. We have
the relative orbital momentum $\ell_{b\bar{b}}$ in the rest frame of the
$b\bar{b}$ system {(which we denote by $B$)}, and also the relative
orbital angular momentum $L$ {of the $BX$ system} in its rest frame which,
of course, is also the centre-of-mass frame of the three body final
state. Assuming that the final state particles are {\em free
  particles}
and well-approximated by undistorted plane waves, Moskal, Wolke, Khoukaz
and Oelert \cite{moskal} have shown that the threshold energy dependence
of the cross section in the 
{$(L, \ell_{b\bar{b}})$ partial wave is given by,}
$$\sigma_{L\ell_{b\bar{b}}} \propto q_{\rm
  max}^{2L+2\ell_{b\bar{b}}+4}\;,$$ where $q_{\rm max}$, the maximum
momentum of the $X$ is given by, $$q_{\rm max}^2 = \lambda(s, 4m_b^2,
m_X^2)/4s.$$ Here, $\sqrt{s}$ is the centre-of-mass energy, and
$\lambda(x,y,z)\equiv x^2+y^2+z^2-2xy-2yz-2xz$. It is straightforward to
show that close to the reaction threshold,\footnote{Of course, the
  $b$-quarks are not produced as free particles and hadronize into
  bottom mesons. The dependence shown in Eq.~(\ref{eq:thresh}) holds as
  long as we are not so close to the threshold that the meson
  binding energy is relevant and not yet so far that the
  non-relativistic approximation used to derive the threshold behaviour
  becomes invalid.}
\be
\sigma_{L\ell_{b\bar{b}}} \propto
  \left(1-\frac{m_X}{\sqrt{s}}-2\frac{m_b}{\sqrt{s}}\right)^{L+\ell_{b\bar{b}}+2}\;.
\label{eq:thresh}
\ee
We stress that in obtaining this we have used non-relativistic
expressions only for the final state wave functions. That the
initial state particles, be they electrons and positrons or even
photons, are relativistic, is irrelevant for our analysis.  We can use
(\ref{eq:thresh}) to obtain the threshold behaviour of the {\em total}
cross section from various initial states since this is dominated by the
lowest values of the exponent $L+\ell_{b\bar{b}}+2$ that is allowed by
symmetries and any other dynamical considerations (see below).

Since the space parity of a Dirac particle-antiparticle fermion pair in
the relative orbital angular momentum state $\ell_{f\bar{f}}$ is given
by $(-1)^{\ell_{f\bar{f}}+1}$, and the corresponding charge conjugation
parity in the total spin state $S_{f\bar{f}}$ is given by
$(-1)^{\ell_{f\bar{f}}+S_{f\bar{f}}}$, we can write the $CP$ parity
($CP$ is conserved by all relevant interactions) of
the $Xb\bar{b}$ system (where, {abusing notation,} we denote $X$ by $h$ ($A$) for $\alpha_b
=0$ ($\pi/2$)) as,
\be
CP_{hb\bar{b}} = (-1)^{L+S_{b\bar{b}}+1}\;, \nonumber \\
CP_{Ab\bar{b}} = (-1)^{L+S_{b\bar{b}}}\;.
\label{eq:cp}
\ee
In order to proceed further, we note that in the process $e^+e^- \to
Xb\bar{b}$, the final state occurs via an intermediate (virtual) $Z$,
and so has total angular momentum $J_{Xb\bar{b}}=1$ and
$CP_{Xb\bar{b}}=1$. $CP$ conservation then implies that, 
\be
(-1)^L(-1)^{S_{b\bar{b}}}=-1 \ {\rm for}  \ X=h, \ {\rm and} \nonumber
(-1)^L(-1)^{S_{b\bar{b}}}=+1 \ {\rm for} \ X=A. \nonumber
\ee
For $L=0$, this then implies that 
\be
S_{b\bar{b}}=1  \ {\rm for} \ X=h, \nonumber \\
S_{b\bar{b}}=0  \ {\rm for} \ X=A. \nonumber
\ee
If, in addition, $\ell_{b\bar{b}}=0$, we conclude that the total angular
momentum of the $b\bar{b}$ system is given by,
\be
J_{b\bar{b}}=1  \ {\rm for} \ X=h, \label{eq:scalar}
\ee
\be
J_{b\bar{b}}=0  \ {\rm for} \ X=A. \label{eq:ps}
\ee
However, $J_{b\bar{b}}=0$ {\em and} $L=0$, together are incompatible
with angular momentum conservation given that the {\em dynamics}
requires that the $b\bar{b}A$ final state comes from a spin-1
$Z^*$. This is not an issue for $b\bar{b}h$ production, as can be seen
from Eq.~(\ref{eq:scalar}). We thus conclude from Eq.~(\ref{eq:thresh})
that close to the production threshold,
\be
\sigma(e^+e^- \to b\bar{b}h) \sim (\sqrt{s}-2m_b-m_h)^2,
\label{eq:scalthresh}
\ee
while
\be
\sigma(e^+e^- \to b\bar{b}A) \sim (\sqrt{s}-2m_b-m_A)^3.
\label{eq:psthresh}
\ee
We stress that $CP$ and angular momentum conservation are, by
themselves, {\em not sufficient} to yield the threshold behaviour of the
cross sections in Eq.~(\ref{eq:scalthresh}) and (\ref{eq:psthresh}). It
is crucial to use the fact that the final state arises from a virtual
$Z^*$, and so has {$J_{Xb\bar{b}}=1$} and $CP=+1$. This is what we had referred to as
``other dynamical considerations'' at the end of the paragraph
containing {Eq.~(\ref{eq:thresh})}.

At first glance, the different
powers in Eq.~(\ref{eq:scalthresh}) and (\ref{eq:psthresh}) appear to be
at odds with the chirality protection mechanism of
Sec.~\ref{sec:chiral}. We have, however, extracted the threshold
behavior of these cross sections from the lengthy expressions 
for the  full calculation of $e^+e^- \to
t\bar{t}+h/A$ production \cite{djouadi} and find that, at the threshold {for the $b \bar{b}+h/A$ case}, 
\be
\sigma(e^+e^- \to b\bar{b}h) \sim (\sqrt{s}-m_h)(\sqrt{s}-2m_b-m_h)^2,
\  {\rm and}  \nonumber  \, \,
\sigma(e^+e^- \to b\bar{b}A) \sim (\sqrt{s}-2m_b-m_A)^3. \nonumber
\ee
We see that the explicit calculation is consistent with the expected
threshold behaviour in Eq.~(\ref{eq:scalthresh}) and
(\ref{eq:psthresh}). Moreover, we  see that, in the chiral limit
$m_b \to 0$, the prefactor $(\sqrt{s}-m_h)$ present in the expression for
$\sigma(e^+e^- \to b\bar{b}h)$ above combines with the second factor to
yield the same behaviour for the $b\bar{b}h$ and $b\bar{b}A$ production,
in agreement with what we would expect from the chirality protection mechanism. 

\begin{figure}[!htp]
\centering
\includegraphics[height=2 in]{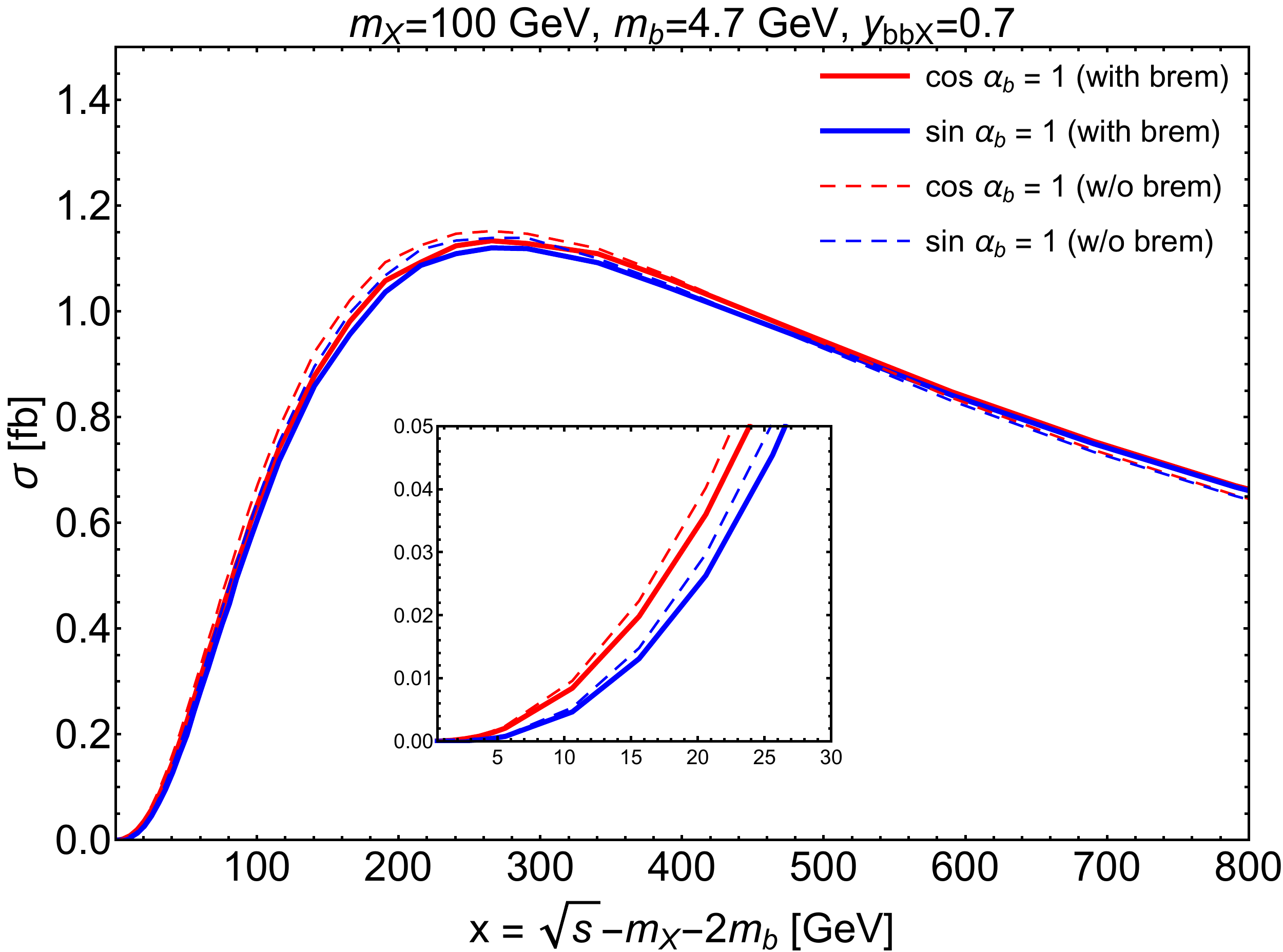}
\includegraphics[height=2 in]{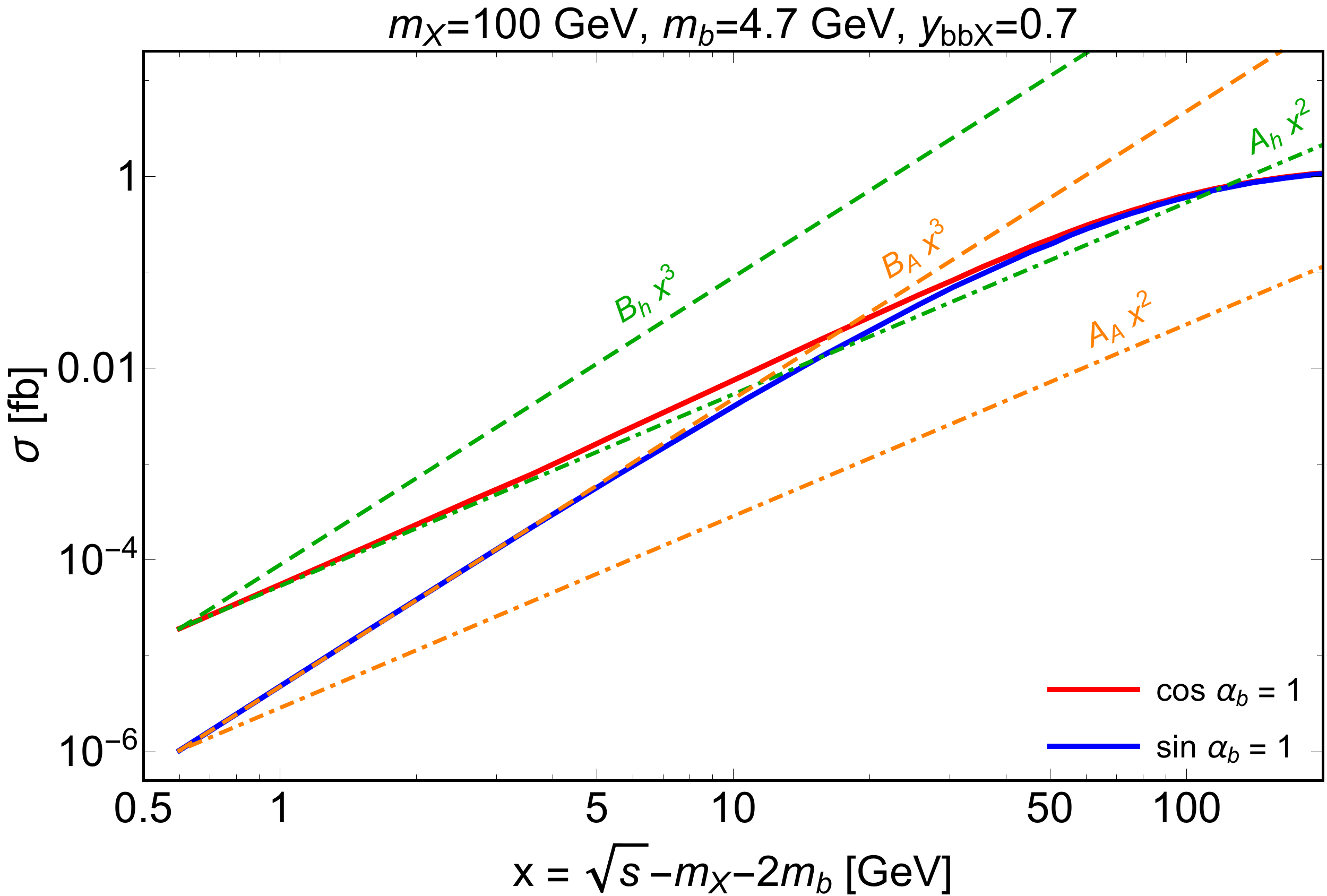}
\caption{Cross section for $ e^+ e^- \rightarrow b \bar{b} X$ process
  (with unpolarized beams) for $\cos \alpha_b = 1$ and $\sin \alpha_b =
  1$, taking $m_X=100$~GeV and $y_{bbX}=0.7$ versus $x=\sqrt{s}-m_X -
  2m_b$, including bremsstrahlung effects on a linear scale [left
    frame]. We also show the same cross section using a log scale [right
    frame] to show the behaviour close to the threshold, $x=0$, along
  with straight lines that illustrate quadratic and cubic rise of the
  cross section. The inset on the left frame shows the same cross section with (solid) and
  without (dashed) bremsstrahlung in linear scale.  
\label{fig:ee2bbX_threshold}}
\end{figure}

To illustrate the threshold behaviour, we show the centre-of-mass energy
dependence of the cross section for $e^+e^- \to b\bar{b}X$ production
for $\alpha_b =0$ (solid red) and $\alpha_b=\pi/2$ (solid blue) in the left
frame of Fig.~\ref{fig:ee2bbX_threshold}, plotting it versus the kinetic
energy release $x=\sqrt{s}-m_X-2m_b$. As before, we fix $m_X$ =100~GeV,
$y_{bbX}=0.7$ and assume that there is no $ZZX$ coupling. We include
bremsstrahlung effects using the Kuraev-Fadin distribution function
\cite{fk} for an electron inside the electron as described in Sec.~2 of
Ref.\cite{krup}. Bremsstrahlung makes only a small shift of the
normalization without any appreciable change in the threshold behaviour
from our expectation as we will see shortly. We do not include any
beamstrahlung effects as these will depend on the as yet undetermined
details of the beam configurations, but we expect this would not have a
qualitative effect on the following discussion. We see from the figure
that the cross section is essentially independent of $\alpha_b$, just as
we would expect from the discussion of
Sec.~\ref{sec:chiral}.\footnote{We emphasize that only the energy
  dependence can be used to distinguish between the two cases because
  the coupling $y_{bbX}$ is completely unknown. Although this is
  completely academic, we note that for the integrated luminosities we
  may anticipate at CEPC or FCC with $\sqrt{s} \sim 240-250$~GeV, {even} the
  normalization difference seen in the figure would be difficult to
  distinguish, since we may expect no more than a few thousand signal
  events per year per interaction point before any acceptance cuts and
  detector efficiencies are included.}

Turning to the right-hand frame, we see that {\em sufficiently close to
  the threshold}, the cross section indeed exhibits the quadratic
[cubic] variation that we anticipated in Eq.~(\ref{eq:scalthresh})
[Eq.~(\ref{eq:psthresh})] for $X=h$ [$X=A$]. Of course, this makes sense
only for $x$ larger than a few GeV so that we can neglect bottom meson
binding effects. Such a study will also require scanning the cross
section close to the reaction threshold. {For $b\bar{b}A$ production, the cubic dependence on $x$ persists for $\lesssim 20$ GeV beyond the
threshold, where the cross section is $\sim 5-20$~ab. In
contrast, we see that the cross section for $b\bar{b}h$ production
begins to deviate from the expected quadratic dependence on $x$ even
$\sim 5$~GeV above threshold,} where the cross section is below 2~ab. It thus
appears that with integrated luminosities of about 10~ab$^{-1}$ ({\it
  i.e.}  $\sim 1$~ab$^{-1}$ per point), may allow confirmation of the
expected threshold behaviour {of $b\bar{b}A$ production,} but significantly
higher integrated luminosities will be needed for the corresponding
study of $b\bar{b}h$ production. We caution that these projections
should only be regarded only as a qualitative indication of what might
be possible at $e^+e^-$ future colliders since we have assumed {a particular value
of $y_{bbX}$,} not included geometric acceptances or experimental
efficiencies, or examined analysis cuts that may be needed to separate
the signal from background.\footnote{We {also} repeat that we have assumed that there is no $XZZ$ coupling,
reminding the reader that in the SM the amplitude containing the $hZZ$
coupling overwhelms $b\bar{b}h$ production.
  }  Our point here is only that
such a threshold analysis may be worthy of further assessment should a
new spin-zero particle that couples to bottom quarks be discovered. A
scan over the $b\bar{b}X$ production threshold may well be the only way
to reveal the spacetime structure of the bottom Yukawa coupling of the
new spin-zero particle.

Before closing this section, we note that we can use similar reasoning
to extract the threshold behaviour of $b\bar{b}X$ production from other
initial states. For instance, for $\gamma\gamma \to b\bar{b}X$, both $C$
and $CP$ are separately conserved, but the final state is not
dynamically constrained to have $CP=1$. We find that
$L=\ell_{b\bar{b}}=0$ is consistent with $C$ conservation if
$S_{b\bar{b}}=0$ since $C_{\gamma\gamma}=+1$. $CP$ conservation then implies
that $b\bar{b}h$ ($b\bar{b}A$) production is allowed as long as the
initial photons are in the  $CP$ odd (even), $J=0$ state. We conclude that,
\be
\sigma(\gamma\gamma \to b\bar{b}X) \sim (\sqrt{s}-2m_b-m_X)^2,
\ X = h, \ A\;. \label{eq:gamgam} \\
\ee
We have verified that the explicit computation yields this threshold
behaviour of the cross sections. 

The threshold behaviour for $b\bar{b}X$ production from $q\bar{q}$
initial states is, of course, identical to that from the
electron-positron initial state, with the virtual gluon playing the role
of the $Z^*$, and constraining the final state to be $CP$
even. Production from the $gg$ initial state is more complicated than
from the $\gamma\gamma$ initial state because now additional amplitudes
are present because of the existence of the three-gluon vertex. In any
case all this is only of academic interest, since at a hadron collider it
will almost certainly be impossible to experimentally study the threshold
behaviour of the production cross section, let alone disentangle the
various subprocesses from each other.

\section{{Top pair production in association with a spin   zero particle}}
\label{sec:top}

\begin{figure}[!htp]
\centering
\includegraphics[height=2 in]{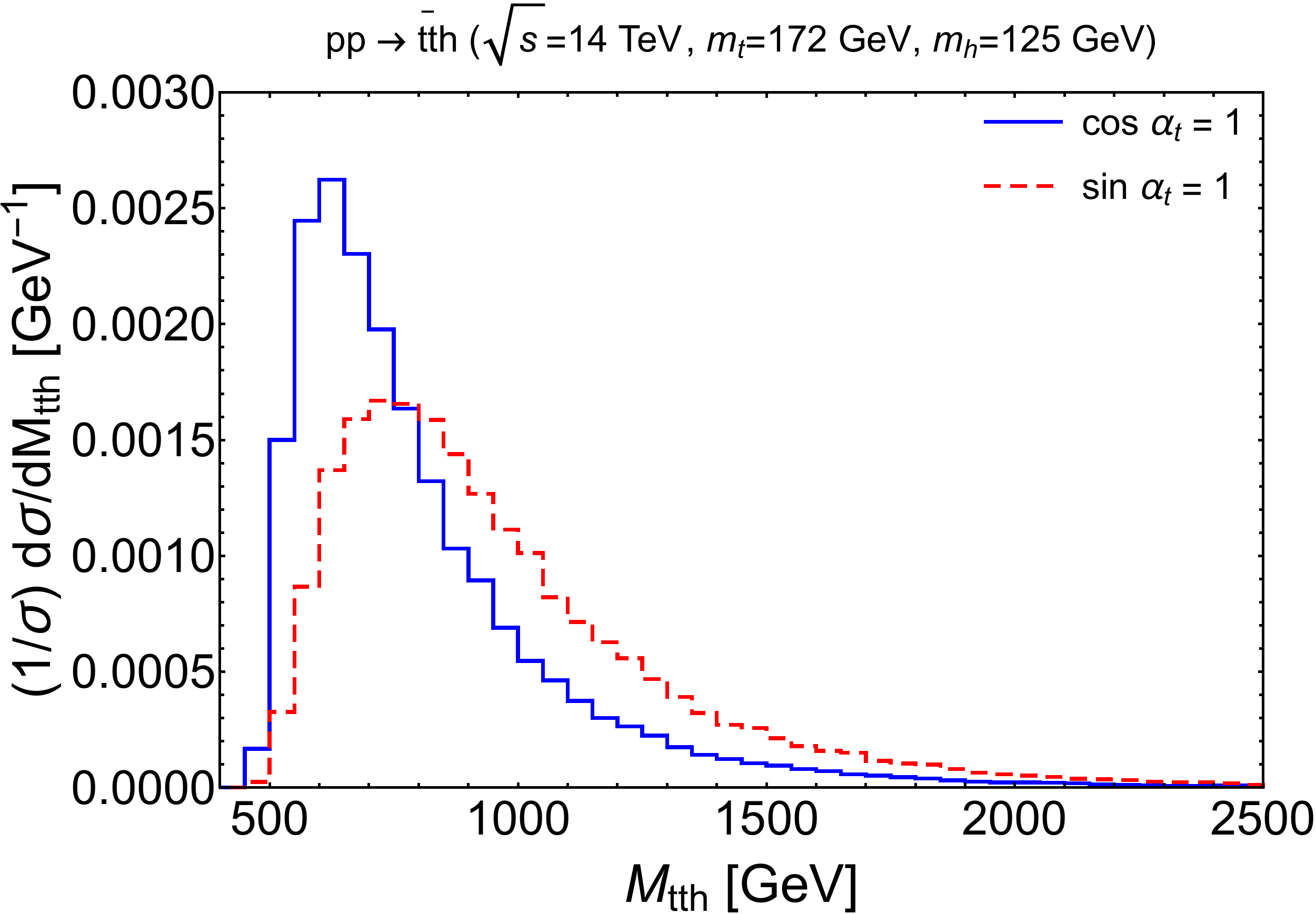}\\
\includegraphics[height=2 in]{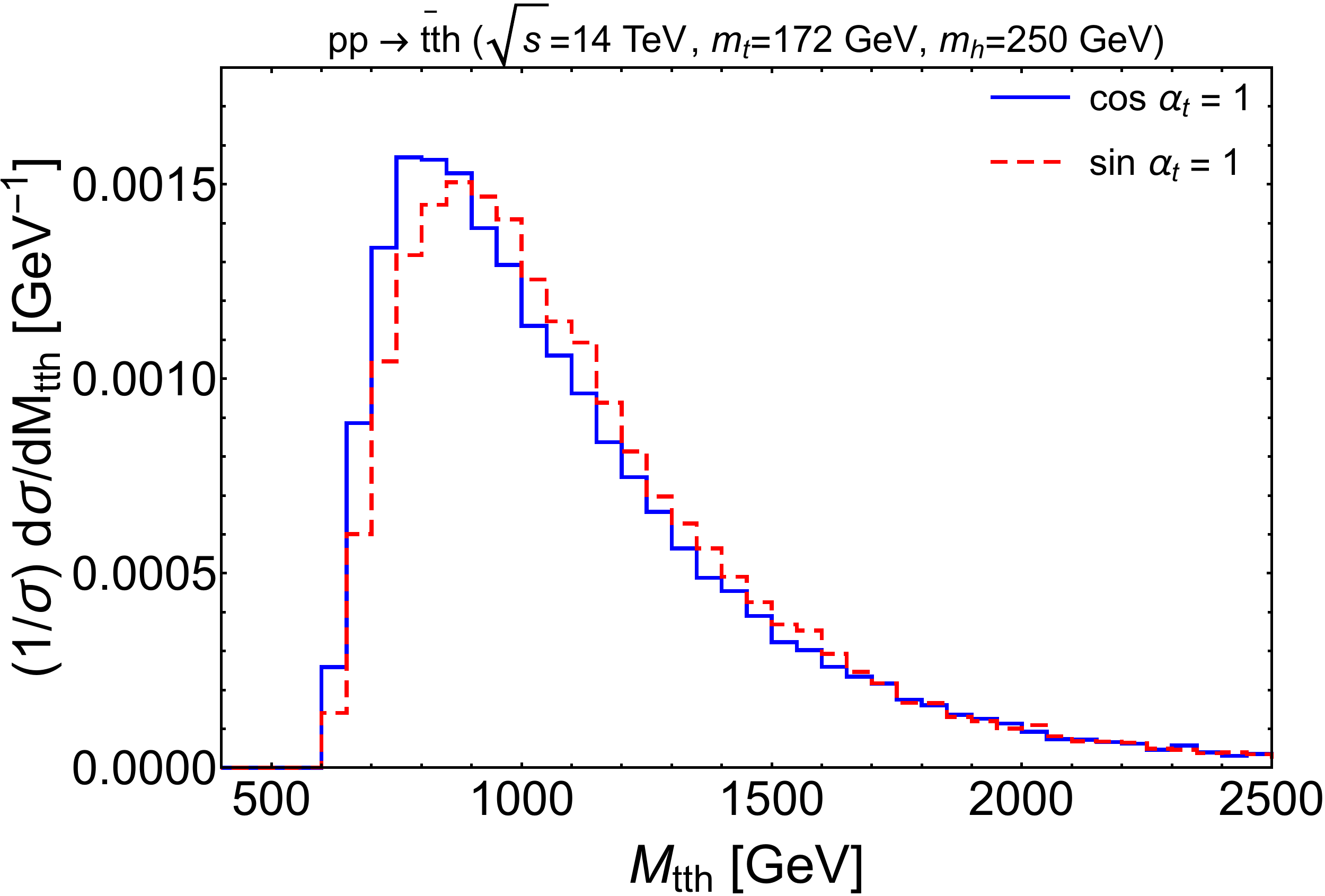}\\
\includegraphics[height=2 in]{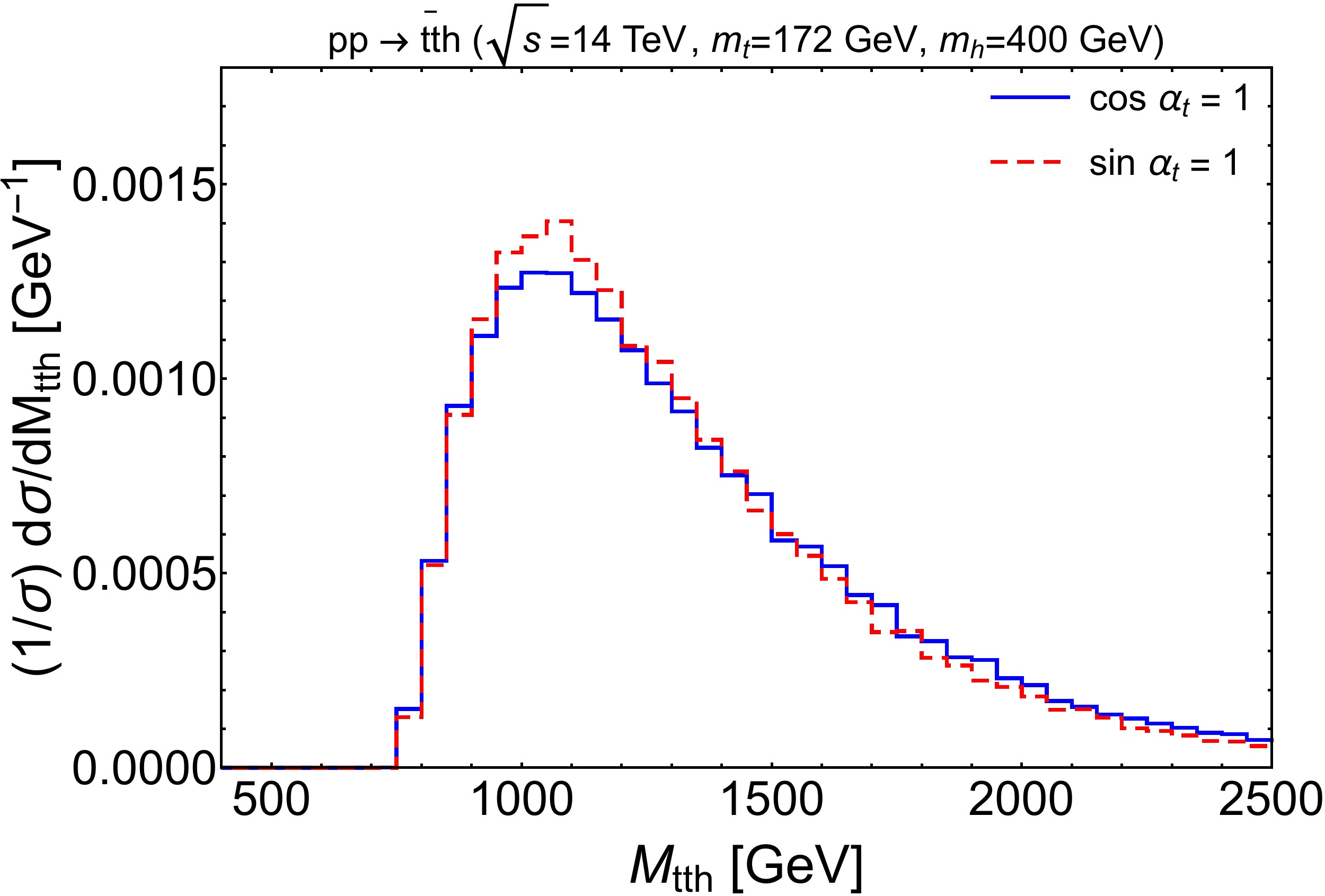}

\caption{{In the top panel we show the invariant mass distributions
  for $t \bar{t}h$ production 
  at the LHC, for $m_h = 125$ GeV. In the two lower panels we show
  invariant mass distributions for $m_h = 250$~GeV and 400 GeV.}
\label{fig:kin_x-ILC_G_fn} }
\end{figure}

We have seen in Sec.~\ref{subsec:sm} that the shape of the
$p_{Th}$ distribution can be used to distinguish between $\alpha_t=0$
and $\alpha_t=\pi/2$ at the LHC\cite{Frederix:2011zi}, while the
chirality protection mechanism precludes the possibility for analogously
pinning down $\alpha_b$.  
{We could equally well have used the $M_{t \bar{t} h}$ distribution to distinguish between $\alpha_t =0$ and $\alpha_t = \pi/2$ as can be seen from the top panel of Fig.~\ref{fig:kin_x-ILC_G_fn}.} What may be
somewhat of a surprise is that the SM Higgs mass is accidentally close to a sweet
spot for distinguishing between the two values of $\alpha_t$.  This may
be seen in the bottom frames of Fig.~\ref{fig:kin_x-ILC_G_fn}, where we
show the $M_{tth}$ distributions for the same two values of $\alpha_t$,
but for $m_h=250$~GeV and 400~GeV.  It is clear that for the larger
values of $m_h$ distinction between $\alpha_t=0$ and $\alpha_t=\pi/2$
becomes much more challenging.  \footnote{We have checked that this
  is also the case for the $p_{Th}$ distribution in $pp\to
  t\bar{t}h$ production at the LHC.} 
  {Although this is somewhat unrelated to the main subject of the paper, we remark that it would be erroneous to infer that one can use kinematic
distributions to extract the spacetime structure of $t\bar{t}X$
couplings, regardless of the mass of $X$. For $m_X \gg m_t$, the distinction between even the extreme cases, $\alpha_t=0$ and $\alpha_t=\pi/2$ vanishes, in accord with the chirality protection argument.}

{Prior to closing this discussion, we mention that we have also checked
that the variation of the $e^+e^- \to t\bar{t}h$ and
$\gamma\gamma \to t\bar{t}h$ cross sections} with the centre of mass
energy $\sqrt{s} = M_{tth}$ also {exhibit} the same qualitative feature:
the energy
dependence of the cross section becomes insensitive to the {spacetime
structure of the top quark Yukawa coupling in Eq.~(\ref{eq:cpvlag}) for large values of $m_h$.}
This
may be understood by noting that the
$t\bar{t}h$ production cross section (or in the case of the LHC, the
hard scattering cross section as a function of parton centre-of mass
energy $\hat{s}$ instead of $s$) takes the form,
\be
\sigma(s,m_t,m_h) = \frac{1}{s}
\left(1-\frac{2m_t}{\sqrt{s}}-\frac{m_h}{\sqrt{s}}\right)^p\times
    G\left(\frac{m_t}{\sqrt{s}}, \frac{m_h}{\sqrt{s}}\right),
\label{eq:dimarg}
\ee 
where we have separated the factor to make explicit the threshold
energy dependence discussed in Sec.~\ref{sec:thresh}.  Any dependence of
the cross section on $\alpha_t$ enters via the dimensionless function
$G$, which encodes the information of the underlying dynamics.  The
chirality protection mechanism implies that the dependence on $\alpha_t$
must vanish if $\frac{m_t}{\sqrt{s}} \to 0$. Since we know that the
cross section in Eq.~(\ref{eq:dimarg}) does not exhibit singular
behaviour as $m_t\to 0$, its sensitivity to $\alpha_t$ is suppressed by
$m_t/\sqrt{s} < m_t/m_h$, accounting for the behaviour seen in
Fig.~\ref{fig:kin_x-ILC_G_fn}, and for the fact that this behaviour
persists for $t\bar{t}h$ production from $\gamma\gamma$ as well as
$e^+e^-$ initial states.\footnote{The argument readily extends to
  the $p_{Th}$ distribution, as long as this distribution is not
  singular behaviour as $p_T \to 0$.}

\section{Summary and Concluding Remarks} \label{sec:concl}

The original motivation for this study was to examine strategies to
determine the spacetime structure of the renormalizable bottom quark
Yukawa interaction with the putative Higgs boson discovered at CERN. Are
these scalar, pseudoscalar or somewhere in between? {We found that adapting strategies that {have} been suggested in the literature for the determination in case of the top quark simply {does} not work for the bottom quark case.}

We traced the underlying reason to the fact that by making suitable
chiral transformations, it is possible to continuously transform between
scalar and pseudoscalar Yukawa interactions with any spin-zero particle,
without affecting its interactions with SM vector bosons. These chiral
transformations alter only the quark mass term. If the quark mass
vanishes (but the Yukawa coupling is held fixed), it is only a matter of
convention whether we call the interaction scalar or pseudoscalar, and
no experimental observable is affected by this change of
description. This is what we dubbed the chirality protection mechanism
in Sec.~\ref{sec:chiral}. In {practice}, any differences between scalar
versus pseudoscalar interactions are suppressed as long as all relevant
sub-process energy scales are large
compared to $m_b$, so that the bottom quarks are relativistic.

Motivated by the fact that a study of $t\bar{t}h$ production had been
shown to allow for promising ways to elucidate the spacetime structure
of the top quark Yukawa coupling, in Sec.~\ref{sec:illus} we examined
$b\bar{b}h$ production both at the LHC as well as at an electron positron
collider, and evaluated several kinematic distributions to quantify the
efficacy of the chirality protection mechanism. We found that, even at
the parton level and without any detector resolution effects, the
differences are too small to allow distinction between scalar and
pseudoscalar Yukawa interactions: more realistic simulations would {reduce
these even further.}

There are several factors that make
the study of the SM bottom quark Yukawa interaction more difficult compared to
that of the top quark. First, the bottom Yukawa coupling is small, so
one suffers from low rates. Second, because the Yukawa coupling is small, the
contribution where the $h$ is radiated from the (virtual) $Z$ dominates
$e^+e^- \to b\bar{b}h$ production and overwhelms the contributions of
interest that involve the bottom quark Yukawa coupling.
\footnote{In addition, the decay products of the top quark retain
  information of the polarization of the parent top, and so provide an
  additional handle. Bottom quark decays {mostly} do not preserve polarization
  information.}  We found that in high energy processes where the bottom
quarks are relativistic, the chirality protection mechanism precludes
the possibility of determining the spacetime structure of the
renormalizable interactions, even for a hypothetical spin-zero particle
with a large coupling to bottom quarks: see Sec.~\ref{subsec:hyp}.  As
discussed in Sec.~\ref{sec:thresh}, close to the reaction threshold for
$e^+e^-\to b\bar{b}X$ production where the bottom quark mass effects are
significant, the chirality protection mechanism ceases to be effective,
leading to a potentially observable difference in the threshold
behaviour for $X=h$ and $X=A$ for integrated luminosities of a few
ab$^{-1}$ per year, and a Yukawa coupling significantly larger than in
the SM.

Are there other ways of differentiating between scalar versus
psuedoscalar couplings of $b$-quarks to spin zero particles? As we have
shown, any differences vanish in the limit of vanishing $b$-quark
mass. One possibility might be the precise determination of the
{rate for $X \rightarrow b \bar{b}$ decay} which is suppressed by $\beta^3$ for $X=h$ but 
by $\beta$ for $X=A$, where
$\beta=\left(1-\frac{4m_b^2}{m_X^2}\right)^{1/2}$. We make no
representation as to whether this might be feasible, but mention it 
more to highlight the difficulty of determining the spacetime structure of
the coupling. 

In summary, our examination of the prospects for determining the
spacetime structure of the Higgs boson Yukawa interaction of bottom quarks in the SM
led us to largely negative conclusions. While the smallness of the SM
Yukawa coupling is an important factor, the real underlying reason for
this is that chirality considerations imply that any difference between
scalar and pseudoscalar interactions vanishes in the limit $m_b\to 0$,
and so is unobservably small in high energy processes where sub-process
energy scales are set by the mass of the spin-zero particle, typically
much larger than $m_b$. Indeed we find that the difference between
scalar and pseudoscalar interactions remain unobservable
even for a hypothetical spin-zero particle with large Yukawa couplings
to bottom quarks.  The energy dependence of the {reaction $e^+e^- \to
b\bar{b}X$ production} close to the
reaction threshold, where the bottom quarks are non-relativistic offers
the best hope for distinguishing between scalar and pseudoscalar
couplings of $X$ to bottom quarks. 

\section*{Acknowledgments}


This work was supported in part by the US Department of Energy, Office
of High Energy Physics Grant No. de-sc0010504.  XT thanks the Centre for High Energy
Physics, Indian Institute of Science Bangalore, where this project
was begun for their hospitality, and also the Infosys Foundation for
financial support that made his visit to Bangalore possible. The work of RG is supported by the Department of Science and Technology, India under Grant No. SR/S2/JCB-64/2007.

%

%
\end{document}